\documentclass[prb,twocolumn,amsmath,superscriptaddress,amssymb,nobibnotes]{revtex4-2}
\usepackage{graphicx}
\usepackage{dcolumn}
\usepackage{bm}
\usepackage{amssymb}
\usepackage{amsmath}
\usepackage{color}
\usepackage[dvipsnames]{xcolor}
\usepackage{placeins}
\usepackage{epstopdf}
\usepackage{multirow}
\usepackage{fontenc}
\usepackage{tabularx}
\usepackage{subfigure}
\usepackage{float}
\usepackage{soul}
\usepackage[normalem]{ulem}
\usepackage[linkcolor=blue,urlcolor=blue,citecolor=blue,colorlinks=true]{hyperref}
\usepackage[parfill]{parskip}    % Activate to begin paragraphs with an empty line rather than an indent

\newcommand{\emp}{{EuMn$_2$P$_2$}}

\begin{document}

\title{Magnetic frustration and weak Mn magnetic ordering in \emp\,} %}\\

\author{Sarah~Krebber}
\email{krebber@physik.uni-frankfurt.de}
\affiliation{Physikalisches Institut, Goethe Universit\"at Frankfurt am Main, Max-von-Laue Straße 1, 60438 Frankfurt, Deutschland}

\author{J\"org~Sichelschmidt}
\affiliation{Max-Planck-Insitut f\"ur
Chemische Physik fester Stoffe, 01187 Dresden, Deutschland}

\author{Pierre~Chailloleau}
\affiliation{Max-Planck-Insitut f\"ur
Chemische Physik fester Stoffe, 01187 Dresden, Deutschland}

\author{Asmaa~El~Mard}
\affiliation{Physikalisches Institut, Goethe Universit\"at Frankfurt am Main, Max-von-Laue Straße 1, 60438 Frankfurt, Deutschland}

\author{Marvin~Kopp}
\affiliation{Physikalisches Institut, Goethe Universit\"at Frankfurt am Main, Max-von-Laue Straße 1, 60438 Frankfurt, Deutschland}

\author{Michael~Baenitz}
\affiliation{Max-Planck-Insitut f\"ur
Chemische Physik fester Stoffe, 01187 Dresden, Deutschland}

\author{Kurt~Kummer}
\affiliation{European Synchrotron Radiation Facility, 38043 Grenoble, France}

\author{Denis~V.~Vyalikh}
\affiliation{Donostia International Physics Center (DIPC), 20018 Donostia-San Sebastian, Spain}
\affiliation{IKERBASQUE, Basque Foundation for Science, 48013 Bilbao, Spain} 

\author{Jens~M\"uller}
\affiliation{Physikalisches Institut, Goethe Universit\"at Frankfurt am Main, Max-von-Laue Straße 1, 60438 Frankfurt, Deutschland}

\author{Cornelius~Krellner}
\affiliation{Physikalisches Institut, Goethe Universit\"at Frankfurt am Main, Max-von-Laue Straße 1, 60438 Frankfurt, Deutschland}

\author{Kristin~Kliemt}
\email{kliemt@physik.uni-frankfurt.de}
\affiliation{Physikalisches Institut, Goethe Universit\"at Frankfurt am Main, Max-von-Laue Straße 1, 60438 Frankfurt, Deutschland}

\date{\today}
\begin{abstract}

%\section{Abstract}{
We report on the electron spin resonance (ESR), heat capacity, magnetization, nuclear magnetic resonance (NMR), magnetic circular and linear dichroism (XMCD, XMLD), as well as the electrical resistivity of EuMn$_{2}$P$_{2}$ single crystals. Antiferromagnetic order of Eu was observed in several quantities at $T^{\rm Eu}_{\rm N}$\,=\,18\,K. The temperature dependencies of ESR linewidth and resonance shift show, when approaching the Eu-ordered state, a divergence towards $T^{\rm Eu}_{\rm N}$, indicating the growing importance of magnetic correlations and the build-up of internal magnetic fields. An additional temperature scale of $\approx 47$\,K has considerable impact on linewidth, resonance field and intensity. This points to the presence of weak Mn-based ordering. 
The observed ESR line is interpreted as an Eu$^{2+}$ resonance, which probes the weak magnetic background of the Mn subsystem. Such picture is suggested by the lineshape which keeps to be Lorentzian across the 47\,K scale and by the ESR intensity which can be described by the same Curie-Weiss temperature above and below 47\,K. In the same temperature range anomalies were observed at 48.5\,K and 51\,K in the heat capacity data as well as a pronounced broadening of the NMR signal of the EuMn$_{2}$P$_{2}$ samples. 
In XMCD and XMLD measurements, this weak magnetic order could not be detected in the same temperature range which might be due to the small magnetic moment, with a potential $c$-component or frustration.

\end{abstract}
\maketitle

\section{Introduction}

In recent times, trigonal antiferromagnetic materials of the Eu$T_2Pn_2$ family ($T=$ Cd, Zn; $Pn=$ P, As, Sb) have become the subject of considerable research interest, as they offer the potential to investigate a range of fascinating physical phenomena including Weyl semimetal states \cite{Ma2019, Wang2019, Su2020}, unusual transport properties such as colossal magnetoresistance \cite{Wang2021, Sunko2023, Usachov2024, Krebber2023, Singh2023, Luo2023, Du2022}, unconventional anomalous Hall effect \cite{Cao2022, Singh2024} or thermoelectric effects \cite{Zhang2008, Zhang2010a}. It has recently been demonstrated that the magnetic order of the Eu$^{2+}$ ions can be altered from AFM to FM in certain instances by means of pressure \cite{Gati2021} or alternative preparation methods \cite{Jo2020, Chen2024_Cd, Chen2024_Zn}. 

The related compound EuMn$_{2}$P$_{2}$ belongs to the same Eu-based 122-systems with the trigonal CaAl$_{2}$Si$_{2}$ crystal structure (space group $P\overline 3m1$ (164)), Fig.~\ref{fig:crystalstructure}. The Eu moment of this compound is located on triangular layers of Eu$^{2+}$ with edge-shared EuP$_{6}$ octahedra, separated by low-carrier density Mn$_{2}$P$_{2}$ blocks. Neutron diffraction on \emp\, revealed A-type AFM Eu order below 16.5\,K with the moments lying in the $a-a$ plane with FM coupling in the plane and AFM coupling along the c direction \cite{Payne2002}. In this neutron diffraction study, no magnetic order of Mn in EuMn$_2$P$_2$ was observed \cite{Payne2002}. This finding was further corroborated in a recent study by heat capacity and susceptibility measurements \cite{Berry2023}. The authors of this study concluded that the long-range Mn magnetic order is suppressed by chemical pressure. Furthermore, the development of short-range Mn magnetic correlations from T $\approx$ 250\,-\,100\,K, interpreted as a precursor to covalent bond formation, was suggested \cite{Berry2023}. This is in contrast to the related compounds EuMn$_{2}$As$_{2}$ and EuMn$_{2}$Sb$_{2}$ where phase transitions attributed to the magnetic order of Mn were visible in the heat capacity at higher temperatures ($T_{\rm N}= 135\,\rm K$, $T_{\rm N}= 128\,\rm K$) \cite{Anand2016, Schellenberg2010}. In trigonal or hexagonal systems, a key ingredient in preventing the onset of magnetic order is geometrical frustration \cite{Han2012, Banerjee2017, Li2021}.

There are a number of electron spin resonanc (ESR) investigations on layered Eu-pnictide systems with \textit{tetragonal} 122-structure whereas the ESR on Eu or Mn containing compounds with \textit{trigonal} 122-structure are rarely reported. An example of the latter is the ESR on powder samples of EuZn$_{2}$(P, As, Sb)$_{2}$ for which ferromagnetic fluctuations were established to dominate the linewidth broadening towards the temperature of antiferromagnetic order \cite{Goryunov2012, Goryunov2014}. \\
It turns out, that the ESR of EuMn$_{2}$P$_{2}$ displays some similarities to that of the Eu-pnictide 122-systems with \textit{tetragonal} structure, such as EuFe$_{2}$As$_{2}$. There one observes a coupling of local Eu$^{2+}$ spins to conduction electrons in the FeAs layers \cite{Dengler2010,Garcia2012,Hemmida2014}. Two transitions are seen in the linewidth: a weak spin-density wave (SDW) transition feature (itinerant iron magnetism) and a divergence for $T$ approaching $T_{N}$ (Eu magnetism). All spectra consist of a single exchange-narrowed Lorentzian line, also for the ESR in EuFe$_{2-x}$Co$_{x}$As$_{2}$ \cite{KrugvonNidda2012} where the linewidth features related to Eu ordering and to the SDW transition depend on the concentration of Co.

The unusual magnetic behavior of \emp\, encouraged us to re-investigate the material in detail in a temperature range which was less studied before using ESR, heat capacity, magnetic susceptibility, NMR, XMCD, XMLD, and resistivity measurements to elucidate the lack of magnetic Mn order with respect to magnetic frustration. Our investigations show that around 50~K a weak magnetic order of Mn sets in, which is explained in detail below.

\section{Experiment}
\subsection{Crystal growth and X-ray diffraction} 
\begin{figure}
\centering
\includegraphics[width=0.9\linewidth]{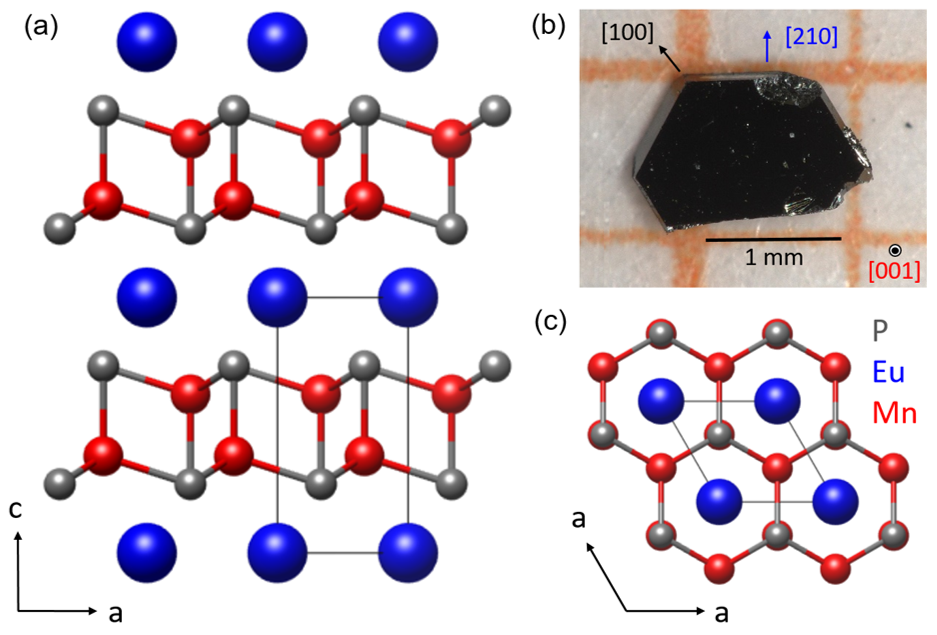}
\caption{\label{fig:crystalstructure} (a) EuMn$_2$P$_2$ crystallizes in the CaAl$_2$Si$_2$ structure type (trigonal space group, $P\overline 3m1$, No. 164). (b) Orientation of a single crystal of EuMn$_2$P$_2$ determined with the Laue method. The surface of the crystal is perpendicular to the $c$-direction. (c) View of the crystal structure along the $c$-direction.}
\end{figure}

Single crystals of \emp~were grown in Sn flux. Pieces of the elements europium (99.99\,\%, Evochem), red phosphorous (99.9999\,\%, Chempur), manganese (99.999\,\%, Evochem) and teardrops of tin (99.9999\,\%, Chempur) were weighed in with a stoichiometry of 14:8:11:265 (Eu:Mn:P:Sn) under inert Ar atmosphere. The elements were then put into a graphite crucible inside an evacuated quartz ampule. The ampule was loaded into a box furnace (Thermconcept), where the mixture was first heated to $450^\circ\text{C}$ and held for $5\,\rm h$. This step ensured that phosphorous could slowly react with the other elements. Subsequently, the mixture was heated to $1200^\circ\text{C}$ and held for $12\,\rm h$. The temperature was then slowly lowered to $970^\circ\text{C}$ with a rate of $2\,\text{K}$/h and afterwards again lowered with a higher cooling rate of $10\,\text{K}$/h to $800^\circ\text{C}$, where the Sn flux was removed by centrifugation. Hexagonal shaped platelet-like single crystals, Fig.~\ref{fig:crystalstructure}(b), were obtained and characterized using powder X-ray diffraction (XRD), energy dispersive X-ray spectroscopy (EDX) and Laue diffraction. XRD analysis yielded  lattice parameters of $a$ = 4.1381\,\AA~ and $c$ = 7.0255\,\AA, which is consistent with literature \cite{Payne2002,Ruehl1979}. XRD data were collected at different temperatures down to 10\,K, confirming that no pronounced structural change in the lattice occurs. The powder diffraction patterns were recorded on a diffractometer by Bruker (D8) with Bragg-Brentano geometry and copper K$_{\alpha}$ radiation. The temperature dependent data was collected on a diffractometer D500 (Siemens) with Bragg-Brentano geometry and copper K$_{\alpha}$ radiation. The orientation of the single crystals was determined by the Laue method using an X-ray equipment “Micro 91” (M\"uller) with a tungsten anode.

\subsection{Electron spin resonance} 
We carried out electron spin resonance (ESR) experiments with a standard continuous-wave spectrometer in a wide temperature range between 5 and 500 K. The ESR setup detects the power $P$ absorbed by the sample from a magnetic microwave field ($\nu\approx9.4$~GHz or 34.1~GHz) as a function of an external, transverse magnetic field $H$. A lock-in technique was used to improve the signal-to-noise ratio which yields the derivative of the resonance signal $dP/dH$.
All spectra were fitted by a Lorentzian shape \cite{Rauch2015} to obtain the parameters linewidth ($\mu_0\Delta H$ being a measure of the spin relaxation), resonance field ($\mu_0H_{\rm res}$, as given by the resonance condition $h\nu = g \mu_B \cdot \mu_0H_{\rm res}$ and being determined by the $g$-value and internal fields), the amplitude, and the ratio of dispersion and absorption contribution $D/A$ (being relevant for electrical conductive samples). The ESR intensity $I_{\rm ESR}$ is deduced from the integrated ESR spectra and is a measure of the local spin susceptibility. The spectra integration was estimated by the amplitude, linewidth, and $D/A$ as exemplified in Ref.~\cite{Gruner2010}. 

\subsection{Heat capacity and magnetic measurements}
Heat capacity (HC) measurements were conducted utilizing the heat capacity option of a Quantum Design Physical Property Measurement System (PPMS). The sample was affixed to a field-calibrated heat capacity puck using Apiezon N grease. In a high vacuum, the sample was subjected to a thermal pulse and its temperature response is recorded in a manner analogous to that employed in a traditional semi-adiabatic relaxation technique. To resolve weak temperature-dependent anomalies large heating pulses were used with typical temperature rises of 6\,K. The data was then analyzed using the single-slope method. The magnetic susceptibility and magnetization measurements were carried out using the vibrating sample magnetometer option (VSM) of the PPMS. The crystal was fixed to the sample holder using GE varnish. The field-dependent data were collected by applying an external magnetic field up to 9 T parallel and perpendicular to the crystallographic c direction of the sample. The analysis of several samples revealed no significant deviations in the heat capacity and magnetization data. 

\subsection{Nuclear magnetic resonance}
Field sweep NMR techniques were applied by use of
the conventional pulsed NMR method on the $^{31}$P nuclei (with nuclear spin $I=1/2$) in the temperature range of $2$~K $\leq$ $T$ $\leq $ $300$~K at fixed frequency of 40\,MHz. The \emp\, powder sample was mixed with paraffin, heated up, shook and then cured to randomize the grains. The field sweep spectra were obtained by integration over the spin echo in the time domain. Powder spectra were modelled with an anisotropic shift tensor. Shift components in the $a-a$ plane and the $c$ direction were determined. 

\subsection{X-ray magnetic circular and linear dicroism}

X-ray absorption spectroscopy measurements were conducted at the ID32 beamline of the ESRF using the high-field magnet end station \cite{brookes2018, kummer2016}. Samples were cleaved in UHV at low temperature immediately prior to the measurements. All spectra were acquired in total electron yield detection mode. Samples were oriented with the [100] and [001] direction in the horizontal plane. X-ray magnetic linear dichroism measurements were performed in normal incidence, i.e. with the incoming X-ray beam along the [001] direction. X-ray magnetic circular dichroism measurements were done in grazing incidence geometry by turning the sample by 60$^\circ$ about the vertical axis with respect to the normal incidence geometry.

\subsection{Resistivity}
Electrical transport measurements on the EuMn$_2$P$_2$ samples were conducted using an AC four-point probe setup and a lock-in amplifier. Low-resistance electrical contacts were achieved by thermally evaporating 200\,nm of gold, with a 7\,nm chromium wetting layer, onto the polished crystal surfaces. Gold wires and silver paste were used for electrical connections. To confirm ohmic behavior, I-V measurements were conducted using a standard DC technique at various temperatures. The resistivity was calculated based on the contact geometry and the dimensions of the a-a plane and c axis.

\begin{figure}[hbtp]
\begin{center}
\includegraphics[width=0.5\textwidth]{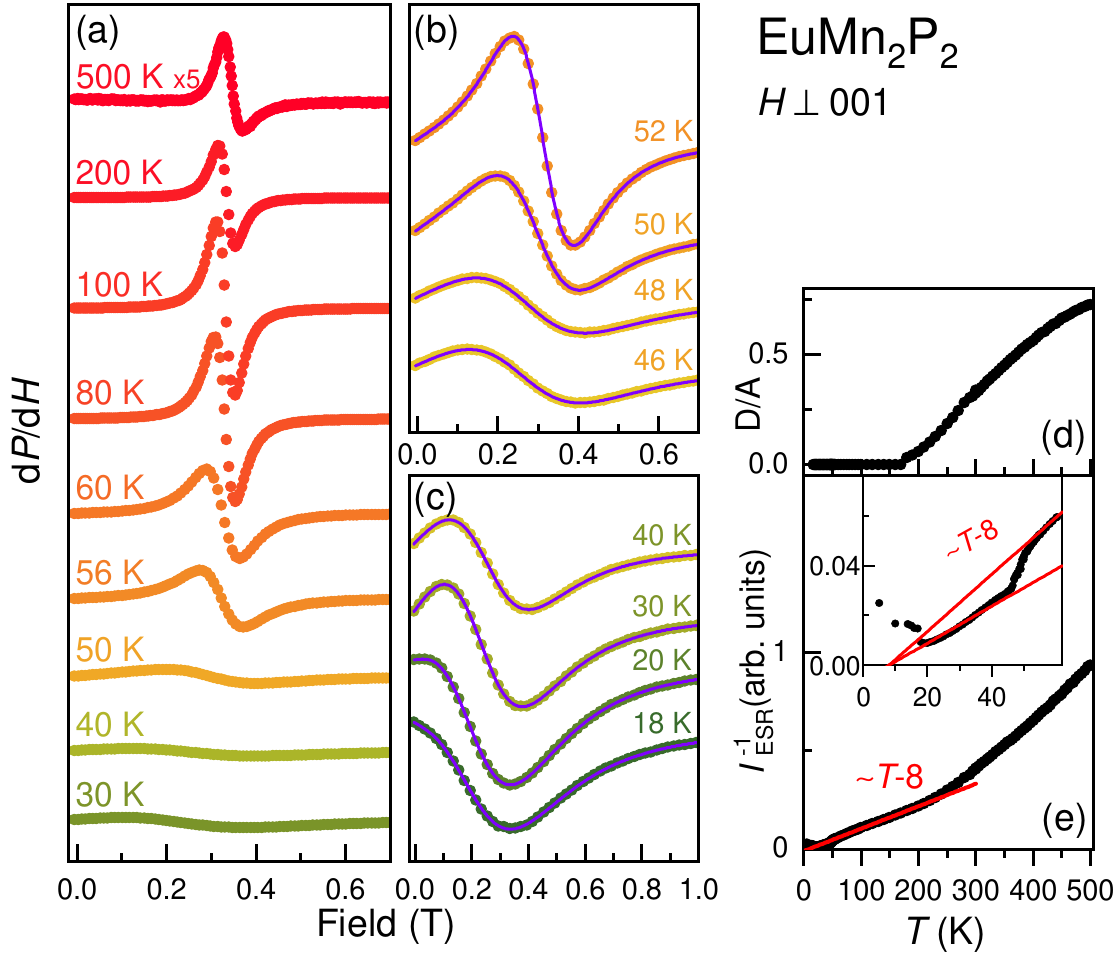}
\end{center}
\caption{Typical ESR spectra (first field-derivative of absorbed power) of single crystalline EuMn$_{2}$P$_{2}$ at X-band (9.4 GHz) and in-plane field direction. (a) Overview for broad temperature region, (b) for region around $T_{\rm M}=47$~K, (c) for region near, but above the magnetic order at $T_{\rm N}$. Solid lines in (b) and (c) denote Lorentzian lineshapes. Temperature dependence of (d) lineshape parameter dispersion-to-absorption, D/A, and (e) reciprocal ESR intensity, enlarged in the inset. Solid lines indicate Curie-Weiss laws $(T-\theta_{\rm W})^{-1}$.}
\label{Fig1ESR}
\end{figure}
%%%%%%%%%%%%%%%%%%%%%%%%%%%%%%%%%%%%%%%%%%%%%%

%%%%%%%%%%%%%%%%%%%%%%%%%%%%%%%%%%%%%%%%%%%%%%%%
\begin{figure}[hbt]
\begin{center}
\includegraphics[width=\columnwidth]{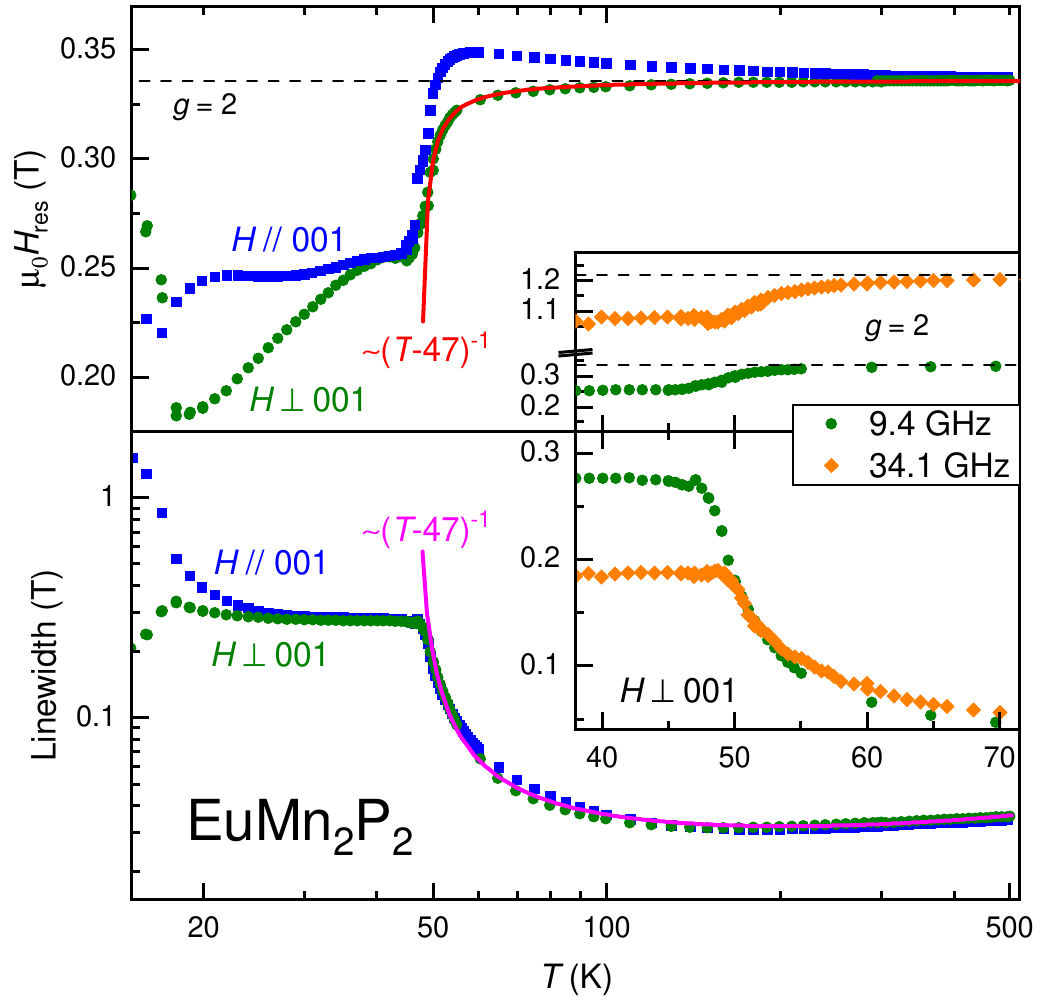}
\end{center}
\caption{ Temperature dependence of resonance field (top frames) and linewidth (bottom frames) for indicated field directions. Solid lines indicate Curie-Weiss behavior. 
}
\label{Fig2ESR}
\end{figure}
%%%%%%%%%%%%%%%%%%%%%%%%%%%%%%%%%%%%%%%%%%%%%%

\section{Results}
\subsection{Electron spin resonance}
The observed ESR signal in EuMn$_{2}$P$_{2}$ consists of a single and well resolved Lorentzian line in the whole temperature range investigated. At 295\,K it corresponds to $g_{\|}=1.984$ ($H\|001$) and $g_{\bot}=2.003$ ($H\bot001$), i.e., $g$-values consistent with an Eu$^{2+}$ resonance as seen in other Eu-pnictides \cite{Dengler2010,Garcia2012,Hemmida2014,KrugvonNidda2012}.
The temperature dependencies of linewidth $\Delta H$ and resonance field $H_{\rm res}$ show clear anomalies not only at $T_{N}$ (Eu-magnetic order) but also at $T_{\rm M}=47$\,K which we relate to Mn spin correlations. 

Fig.~\ref{Fig1ESR}(a-c) shows ESR spectra at selected temperatures in the paramagnetic region. A Lorentzian shape, solid line, describes the spectra perfectly well, in particular in the region around $T_{\rm M}=47$~K,  Fig.~\ref{Fig1ESR}(b), and near Eu-magnetic order at $T_{\rm N}$, Fig.~\ref{Fig1ESR}(c).  
As will be shown below, an anomalous behavior of linewidth and resonance field appears around $T_{\rm M}$. For $T<T_{\rm N}$ double line spectra appear (not shown) as being characteristic for the magnetic resonance of ordered spin systems \cite{sichelschmidt18a}.
A finite dispersion contribution to the lineshape, $D/A>0$, appears around $T=150$~K, see Fig.~\ref{Fig1ESR}(d). Above this temperature, the microwave penetration depth falls below the sample size as a consequence of an increasing sample conductivity. 
%At $T=300$~K, $\rho \simeq 0.7~\Omega \rm{cm}$ and with $\delta(T)\propto\sqrt{\rho(T)/\nu}$ $\delta(300~\rm{K})=0.4$~mm which is in the order of the sample thickness ($\simeq 1$~mm).
At $T=500$~K, $D/A\simeq0.7$ which is reflected in a clear asymmetry in the lineshape as can be seen in the top spectrum of Fig.~\ref{Fig1ESR}(a).

The ESR intensity as deduced from the integrated ESR spectra \cite{Gruner2010} is depicted in Fig.~\ref{Fig1ESR}(e). A clear kink is visible around $T_{\rm M}=47$~K above and below which $I_{\rm ESR}$ is characterized by a Curie-Weiss behavior with $\theta_{\rm W}=+8$~K. 
Around 200~K, an upward bending of $I^{-1}_{\rm ESR}(T)$ is observed. This is most probably due to a microwave skin depth ($\delta$) reduction as indicated by a strong increase of the $D/A$ ratio, see Fig.~\ref{Fig1ESR}(d). In this case, the approximation in calculating $I_{\rm ESR}(T)$ (Ref.\cite{Gruner2010}) is not valid.

The temperature dependencies of the linewidth and the resonance field are depicted in Fig.~\ref{Fig2ESR}. Both quantities show remarkable shoulder-form anomalies around $T_{\rm M}=47$~K. By decreasing the temperature across $T_{\rm M}$ the line broadening abruptly stops, whereas at the same time the resonance field strongly decreases. These features depend on the applied field, as shown in the inset of Fig.~\ref{Fig2ESR}. The position of the shoulders shift to higher temperatures when shifting the resonance to higher fields by changing the microwave frequency from 9.4~GHz to 34~GHz. This behavior is also consistently observed in specific heat data.    
Towards high temperatures, $T\rightarrow 500$~K, the resonance field merges the field corresponding to $g=2$ (dashed lines in Fig.~\ref{Fig2ESR}) being a typical value for Eu$^{2+}$ while the linewidth contains a linear contribution, indicating a Korringa relaxation, and a tendency to saturation. The latter indicates exchange narrowing according to the Kubo-Tomita mechanism \cite{Kubo1954,Huber1999}. 

The approach from above towards the magnetic order at $T_{\rm N}$ influences the resonance field to shift in opposite directions for the two field directions applied. This is quite similar to the shift reported for EuFe$_{2-x}$Co$_{x}$As$_{2}$ where it indicates increasing internal fields near the magnetic order of the Eu spins \cite{KrugvonNidda2012}. For EuMn$_{2}$P$_{2}$, the reduction of the resonance field near $T_{\rm N}$ is strongest for the in-plane field direction, which is consistent with in-plane ferromagnetically correlated Eu$^{2+}$ spins. Approaching $T_{\rm M}$ from above, the in-plane resonance field reflects the evolution of the in-plane magnetization, showing a Curie-Weiss dependence $\propto (T-47)^{-1}$ (see red solid line in Fig.~\ref{Fig2ESR}).

For the linewidth, when decreasing the temperature towards $T_{\rm N}$, one typically expects a broadening due to increasing spin correlations. Above $T_{\rm M}=47$\,K, however, this mechanism obviously is superimposed by the effect of spin correlations related to the anomaly at $T_{\rm M}$. These correlations can be described by a $(T-47)^{-1}$ behavior in the linewidth (see pink solid line in Fig.~\ref{Fig2ESR}). Thus, the broadening shows a critical behavior with a critical exponent of $-1$ which is observed in systems with spin fluctuations in a 3D Heisenberg ferromagnet \cite{benner90a}.

\subsection{Heat capacity}

\begin{figure}
\centering
\includegraphics[width=1.0\linewidth]{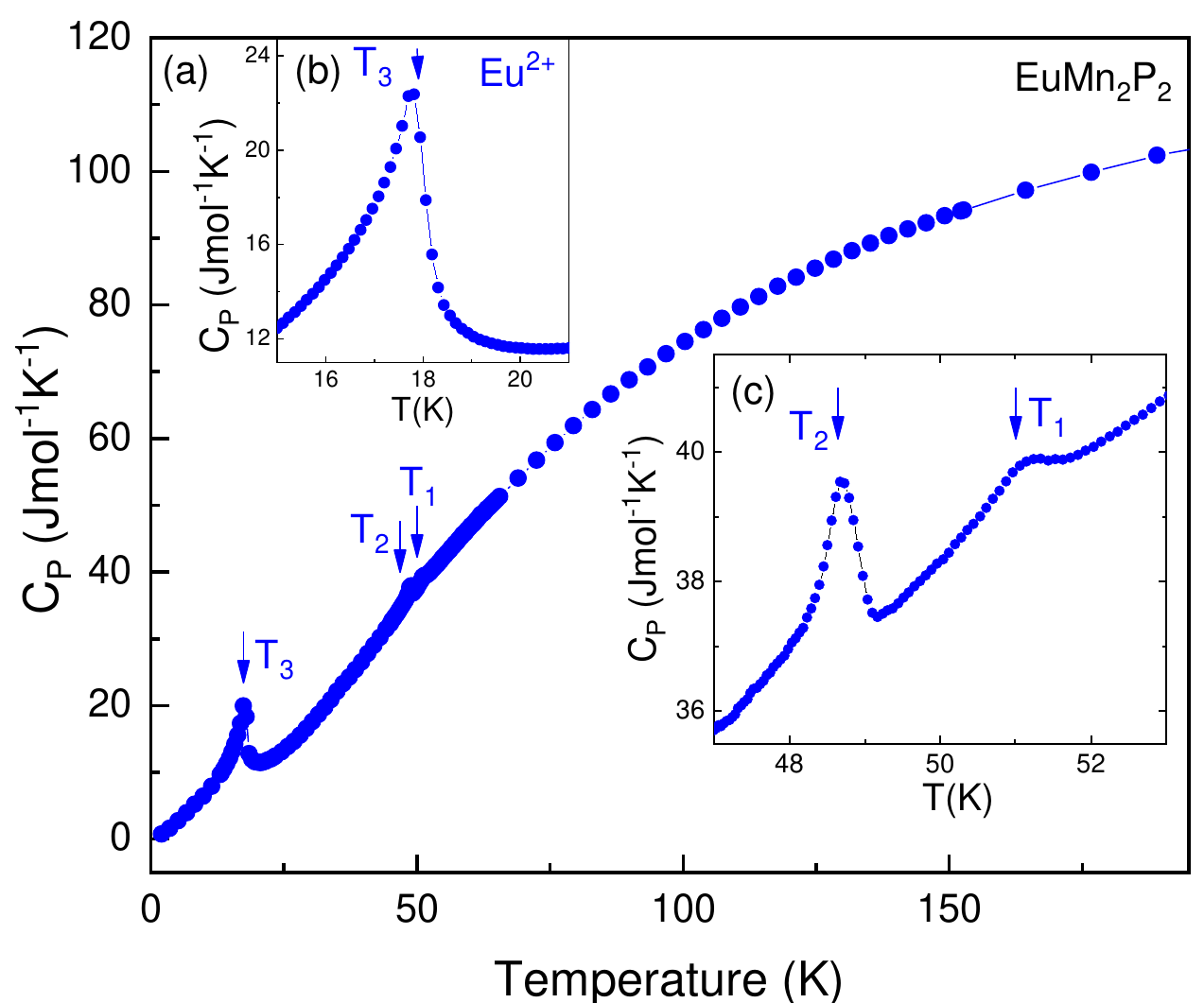}
\caption{\label{fig:AE105_HC_fein} EuMn$_2$P$_2$ (a) Heat capacity at $\mu_0H$ = 0\,T as a function of temperature (blue). (b) Enlarged view of the data in the vicinity of the magnetic transition of Eu. (c) Enlarged view of the data in the vicinity of the magnetic transition of T$_2$ and T$_3$.}
\end{figure}

Fig.~\ref{fig:AE105_HC_fein}(a) shows the zero-field heat capacity of EuMn$_2$P$_2$ between 2\,K and 200\,K. At low temperatures, a sharp $\lambda$-type anomaly appears with a peak at T$_{3}$ = $T^{\rm Eu}_{\rm N}$ = $18\,\rm K$, Fig.~\ref{fig:AE105_HC_fein}(b), which can be attributed to the Eu$^{2+}$ AFM-order. Fig.~\ref{fig:AE105_HC_field_Eu} \cite{Krebber2025_SI} shows the field dependence of this transition with the field applied along the out-of-plane direction, $H\parallel 001$, between 11\,K and 19\,K. 
An increase in the external magnetic field results in a shift of the transition temperature T$_3$ towards lower temperatures, which is consistent with an AFM ordering of the Eu$^{2+}$ spins. The peak is completely suppressed for fields exceeding $\mu_0H=5\,\rm T$.
In addition, as shown in Fig.~\ref{fig:AE105_HC_fein}(a,c), two small anomalies at $T_2=48.5\,\rm K$ and $T_1=51\,\rm K$ were observed in zero field after precise measurements in the corresponding temperature range where the significant change in the resonance field and linewidth in the ESR data occurs. Surprisingly, in addition to the anomaly at $T_2$, a further anomaly was detected in the heat capacity at around $T_1$, which was not resolved in the ESR data. 

\begin{figure}
\centering
\includegraphics[width=0.9\linewidth]{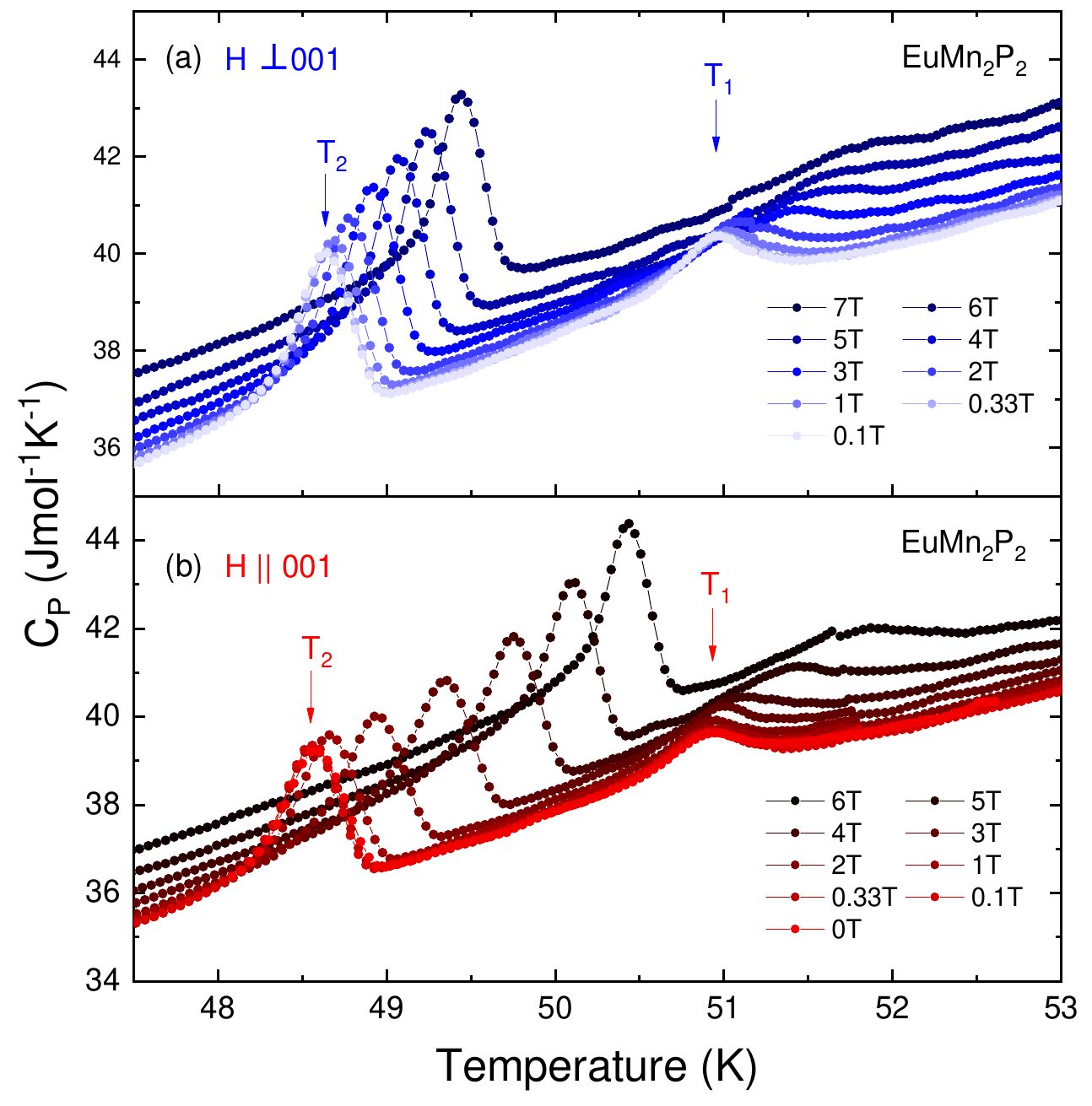}
\caption{\label{fig:AE105_HC_field_Mn} Heat capacity of EuMn$_2$P$_2$ measured in different magnetic fields (a) $H\perp001$ and (b) $H\parallel001$ between 47\,K and 53\,K as a function of temperature. The magnetic transitions at $T_1$ and $T_2$ are assigned to a weak ordering of Mn.  The peak at  $T_1$ shifts to higher temperatures and broadens in field. In contrast, the peak at $T_2$ sharpens in field, shifts to higher temperatures with increasing field, and shows an anisotropy for the in-plane and out-of-plane directions.}
\end{figure}

Fig.~\ref{fig:AE105_HC_field_Mn} shows the field-dependent heat capacity close to the $T_1$ and $T_2$ transitions for $H\perp 001$ and  $H\parallel 001$. 
The T$_1$ transition shifts to higher temperatures, and the shift seems to be almost isotropic for field aligned along the two crystallographic directions. In addition, for both field directions the peak at T$_1$ gets broader with increasing field. In contrast, the peak at $T_{2}$ sharpens with increasing field for both directions and presents an anisotropic shift towards higher temperatures. With field aligned along the $[001]$ direction, the shift to higher temperatures is larger compared to the in-plane direction. Our finding is in contrast to the heat capacity results discussed in Ref.~\cite{Berry2023} where only one Eu$^{2+}$ AFM-transition at T$_3$ was detected, probably due to the lower T-resolution of those measurements.

Furthermore, to rule out additional transitions the zero-field heat capacity was measured precisely at  higher temperatures (see Fig.~\ref{fig:HC_hightemp} \cite{Krebber2025_SI}) comprising the $T$ region around $T=150\,\rm K$  where a reoccurring change in slope in the resistivity data was observed as discussed later. However, no further transitions were found between $T=50-200\,\rm K$. 

\subsection{Magnetic susceptibility and magnetization}

\begin{figure}
\centering
\includegraphics[width=1\linewidth]{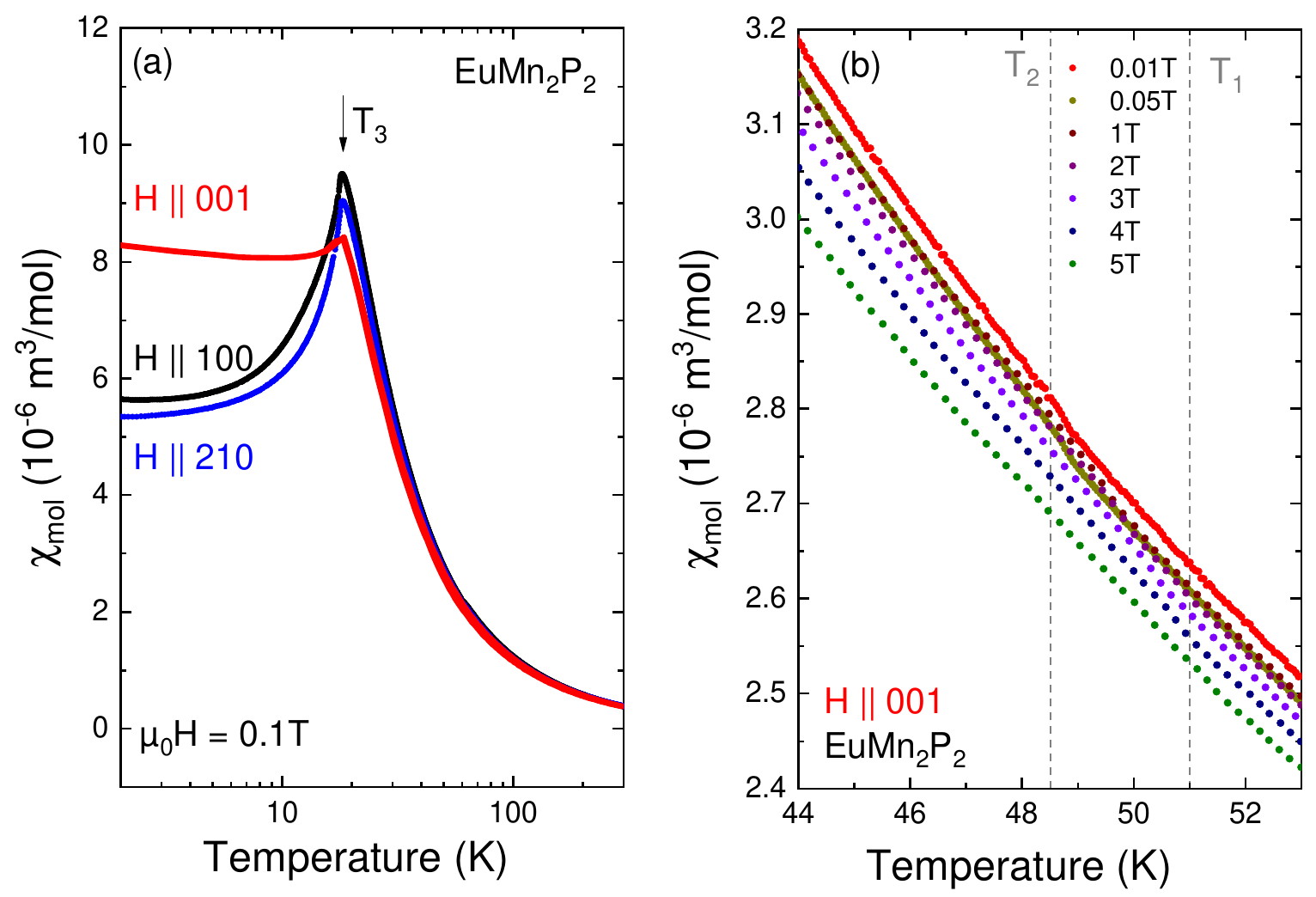}
\caption{\label{fig:MvT_01T_2.pdf} (a) Molar magnetic susceptibility $\chi_{\rm mol}(T)$ at $\mu_0H=0.1\,\rm T$ for $H\parallel 100$, $H\parallel 210$ and $H\parallel 001$. (b) $\chi_{\rm mol}(T), H\parallel 001$: The two magnetic transitions at $T_1$ and $T_2$ are hardly visible in the magnetic susceptibility. Dashed lines indicating the transition temperatures from the zero field HC are guides to the eyes.}
\end{figure}

The magnetic behavior of EuMn$_2$P$_2$ was analyzed by measuring the magnetic susceptibility and magnetization at different magnetic fields. 
Fig.~\ref{fig:MvT_01T_2.pdf}(a) shows the molar magnetic susceptibility $\chi(T)$ of EuMn$_2$P$_2$ at $\mu_0H=0.1\,\rm T$ for the in-plane ($H\parallel210$, $H\parallel100$) and the out-of-plane directions ($H\parallel001$). The data exhibit a clear magnetic transition at  $T_{3}=18\,\rm K$ arising from AFM ordering of the Eu$^{+2}$ spins. Furthermore, an anisotropy of the in-plane and out-of-plane directions is seen below T$_3$. This is in contrast to the data in \cite{Berry2023} where a different kind of anisotropy was detected.
The temperature-dependent magnetic susceptibility with various fields applied along the in-plane and out-of-plane directions is depicted in Fig.~\ref{fig:AE114_Chi_field001} \cite{Krebber2025_SI}.
The AFM nature of the Eu$^{2+}$-transition is confirmed by the shift of $T_{3}$ with increasing field towards lower temperatures for both field directions. The drop in the susceptibility at 3.7\,K at low fields results from superconducting Sn inclusions inside the sample. The Sn content can vary significantly between samples, with values ranging from (1-5)\%. 

Fig.~\ref{fig:AE114_Chi-1} \cite{Krebber2025_SI} shows the reciprocal magnetic susceptibility $\chi^{-1}(T)$ as a function of temperature for the in-plane and out-of-plane directions in the range of 2-300\,K measured with an applied magnetic field of $\mu_0H$ = 1\,T. The data was fitted using the Curie-Weiss law $\chi^{-1}=(T-\Theta_W)/C$ with a Curie constant $C=110\,\text{m$^3$K/mol},$ between 150\,K and 290\,K, which leads to an effective moment of $\mu_{\text{eff}}$ = (8.4\,$\pm$\,0.1)\,$\mu_{\text{B}}$ per formula unit and a Weiss temperature $\Theta_W$ = (7$\pm$\,1)\,K for field along the $[001]$ direction and of $\mu_{\text{eff}}$ = (8.4\,$\pm$\,0.1)\,$\mu_{\text{B}}$ and $\Theta_W$ = (10$\pm$\,1)\,K for field along an in-plane direction in agreement with \cite{Berry2023}. The experimental effective moment is therefore slightly larger compared to the free-ion value of 7.94\,$\mu_{\text{B}}$ for divalent europium, which may be an indication for further contributions to the magnetic susceptibility from the Mn spins. Below $\approx 130~\rm K$, we observe a minor deviation from Curie-Weiss behavior which would be consistent with a development of short-range Mn magnetic correlations which was already discussed in Ref.~\cite{Berry2023}.\\
Fig.~\ref{fig:MvT_01T_2.pdf}(b) shows $\chi(T)$ in the vicinity of the observed $T_1$ and $T_2$ anomalies as detected in the heat capacity (Fig.~\ref{fig:AE105_HC_field_Mn}). However, no clear feature was seen around the $T_1$ and $T_2$ transition temperatures in $\chi(T)$. Mainly, a slight change in slope is observed. The same temperature range was furthermore measured in field-cooled (FC) and zero-field cooled (ZFC) mode for both field directions (Fig.~\ref{fig:AE109_ChivT_FC-ZFC} \cite{Krebber2025_SI}). %Similar to Fig.~\ref{fig:AE114_Chi_field_50K},  
Also here, no clear feature was observed around the $T_1$ and $T_2$ transitions. 

Fig.~\ref{fig:AE114_MvH_Vgl4K}(a) \cite{Krebber2025_SI} shows the field-dependent magnetic moment $M(H)$ per Eu for field aligned along the different crystallographic orientations at 4\,K. The saturation magnetization value is evaluated as 7$\mu_\text{B}$ at $\mu_0$H = 8\,T and 4\,K. The sample mass was corrected according to the amount of Sn inclusions inside the sample, as estimated from the drop of $\chi(T)$ below the superconducting transition of Sn (Fig.~\ref{fig:AE114_Chi_field001} \cite{Krebber2025_SI}). The in-plane directions behave nearly isotropic and make the magnetic easy plane of the system with a critical field of $\mu_0H^{100}_c= 4.7\,\rm T$. In the out-of-plane direction, the saturation magnetization is reached at a critical field of $\mu_0H^{001}_c= 5.5\,\rm T$. The low-field behavior of the moment per Eu$^{2+}$ for the three main symmetry directions, presented as $M/\mu_0H$ versus $\mu_0H$ is shown in Fig.~\ref{fig:AE114_MdurchH_4K}(b) \cite{Krebber2025_SI}. 
From these data, it is possible to deduce information about the alignment of the Eu magnetic moments in the ground state of the material. Our result hints to moment alignment close to the $[100]$ direction and is consistent with neutron diffraction which revealed A-type AFM Eu order with moments lying in the $a-a$ plane at 1.4\,K with FM coupling in the plane and AFM coupling of the nearest-neighbour spins along the $c$ axis \cite{Payne2002}. 

A hysteresis opens between  $\mu_0H$ = 2.5\,T and 4\,T for both field directions, which becomes larger at 10~K, Fig.~\ref{fig:MvH_10K.pdf}(c) \cite{Krebber2025_SI}. For the in-plane direction, this hysteresis is more pronounced. 
Figs.~\ref{fig:AE114_MvH_210} and \ref{fig:AE114_MvH_50K} \cite{Krebber2025_SI} show the $M(H)$ data for different higher temperatures, especially those collected in the temperature range of the $T_1$ and $T_2$ transitions  with field along the $[001]$ direction. The $M(H)$ data are almost linear in $H$ at each temperature, and no significant change in slope was detected at the transition temperatures.

\subsection{Nuclear magnetic resonance}

\begin{figure}
\centering
\includegraphics[width=\columnwidth]{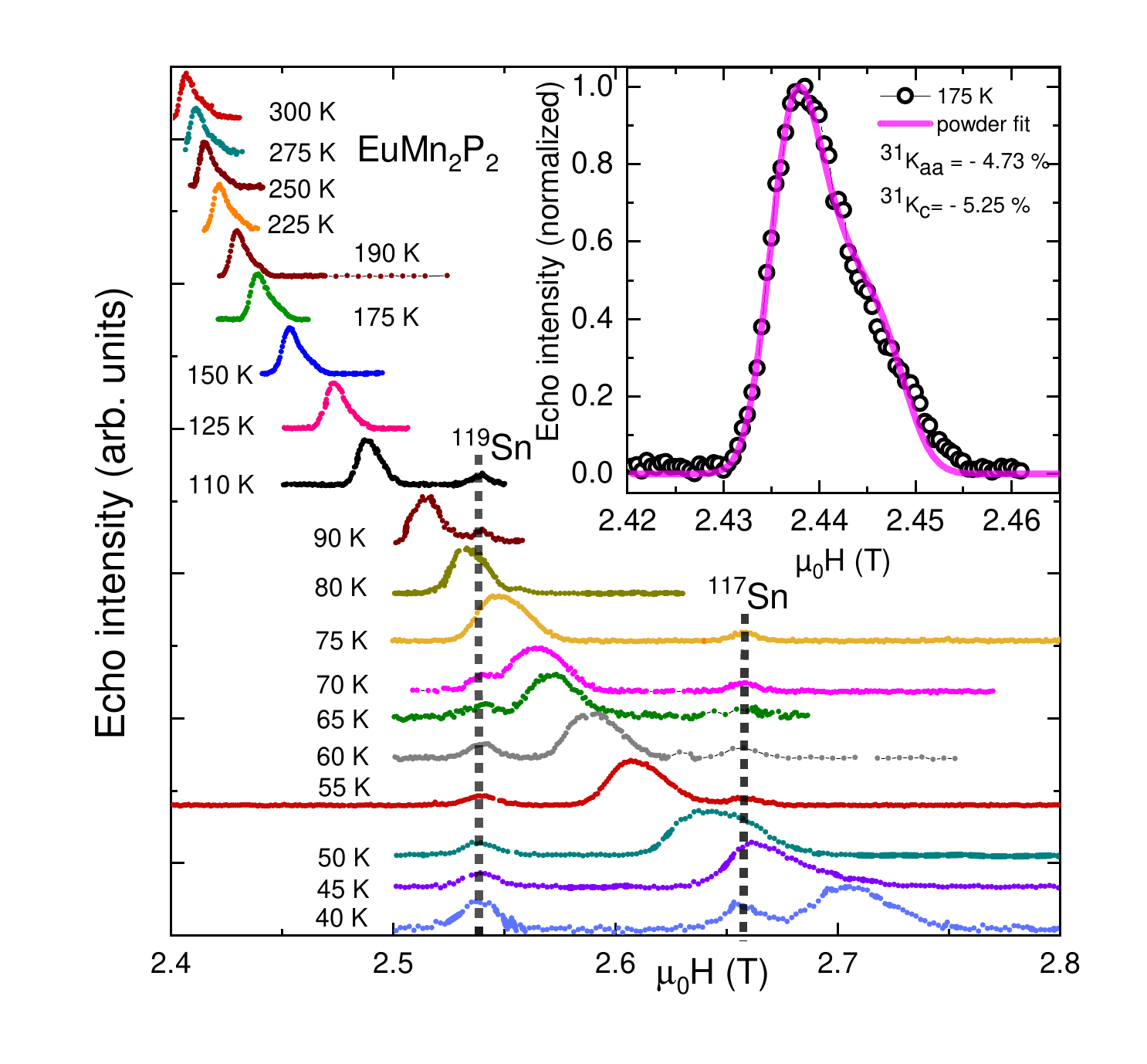}
\caption{\label{fig:NMRFig1} $^{31}$P field sweep NMR spectra taken at 40\,MHz at various temperatures as indicated. The spectra were normalized to the maximum signal of the $^{31}$P NMR line (of each individual spectra) and shifted by a constant value in the vertical direction for better clarity.  The vertical dotted lines indicates the Larmor fields for
$^{119,117}$Sn. The $^{31}$P NMR line shows a sizable shift and broadening towards lower temperatures, whereas the $^{119,117}$Sn lines does not show any temperature dependent shift, as expected for a Sn containing non-magnetic impurity. The inset shows the $^{31}$P NMR line at 175 K together with a theoretical powder simulation (solid line).}
\end{figure}

\begin{figure}
\centering
\includegraphics[width=\columnwidth]{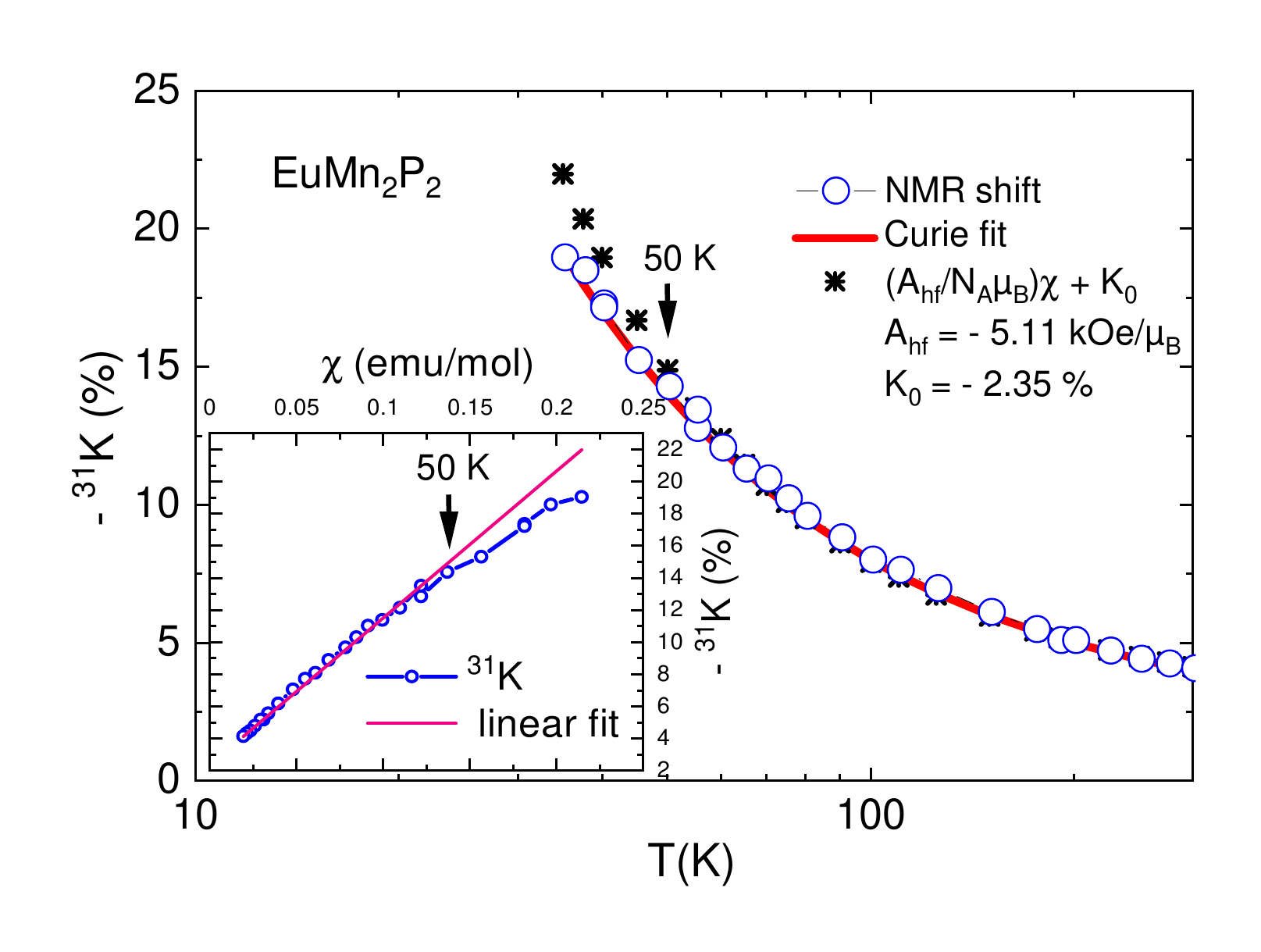}
\caption{\label{fig:NMRFig2} Temperature dependence of the $^{31}$P isotropic shift $^{31}K_\mathrm{iso}$ in EuMn$_2$P$_2$ estimated from 40 MHz field sweep NMR. The curved line indicates a Curie fit to the data (see text). The star symbols correspond to shift values calculated from the determined hyperfine coupling constant. The inset shows the $K$--$\chi$ plot together with a linear fit above 50 K. Note that here the susceptibility at 2.45\,T was used.}
\end{figure}

The NMR spectra of a  polycrystalline powder sample of EuMn$_2$P$_2$ were investigated at a fixed frequency $\nu = 40$ MHz in the temperature range between 300\,K and 40\,K. The results are shown in Fig.~\ref{fig:NMRFig1}. The spectra show a characteristic pattern of randomly oriented powder sample in a layered environment. The lineshape was simulated to extract the shift values according to the $a-a$ plane $^{31}K_{aa}$ and c direction $^{31}K_{c}$, respectively. The inset shows a simulation together with the data for 175\,K. The $^{31}$P NMR line shows a negative total shift with a slightly larger value perpendicular to the $a-a$ plane. The shift values increase towards lower temperatures, and the anisotropy becomes less visible below approximately 125\,K. Towards lower temperatures, a broadening of the line could be seen. Below 40\,K, a measurement of the $^{31}$P NMR signal becomes very complicated due to the fact that the relaxation becomes very fast and the signal wipes out. Before we discuss the wipe out effect, we discuss the temperature dependent shift. 

The transferred hyperfine field at the P ligand site originates mainly from the Eu and Mn based magnetism. As usual we have a combination of mechanisms for the transfer. Conduction electron polarization, core polarization through d electrons and superexchange are usually discussed to originate the hyperfine field at the ligand site. The hyperfine field causes a shift of the NMR line with respect to the nonmagnetic reference compound. Usually the temperature dependent part of the shift is related to the bulk susceptibility via the hyperfine coupling constant. In addition, a small temperature independent part is frequently found which could be related either to conduction electrons (positive) or orbital contributions (negative). 

The total Knight shift can be described by a sum of two contributions:
\begin{equation}
 \label{eq:Kshift}
  \mathcal{K} = \mathcal{K}_{0}+\mathcal{K}_{\rm 4f/3d}(T).
\end{equation}
The temperature dependent shift contribution is related to the bulk susceptibility, giving:
\begin{equation}
  \label{eq:Kchi}
 \mathcal{K}_{\rm 4f/3d}(T) = \frac{A_{\rm hf}}{N_{A}\mu_{B}}\chi(T),
\end{equation}
where $A_{\rm hf}$ is the hyperfine coupling constant, $N_{A}$ the Avogadro number and $\mu_{B}$ the Bohr magneton \cite{Carter1977}.
We used phosphorous acid H$_{3}$PO$_{4}$ with $^{31}$K = 0 as a reference compound for the absolute shift determination. Fig.~\ref{fig:NMRFig2} shows the isotropic shift given by $^{31}K_\mathrm{iso} = (2K_{a,b}+ K_c)/3$ as a function of temperature.  The "Clogston-Jaccarin" plot ($K$--$\chi$ plot, see Fig.~\ref{fig:NMRFig2} inset) relates the NMR shift $^{31}K_\mathrm{iso}$ to the bulk susceptibility of the polycrystalline sample. We obtain the hyperfine coupling constant of $A_\mathrm{hf} = -5.11\,$kOe/$\mu_B$ from the observed linearity of this plot and a $K_0$ value of -2.35\%.

\begin{figure}
\centering
\includegraphics[width=\columnwidth]{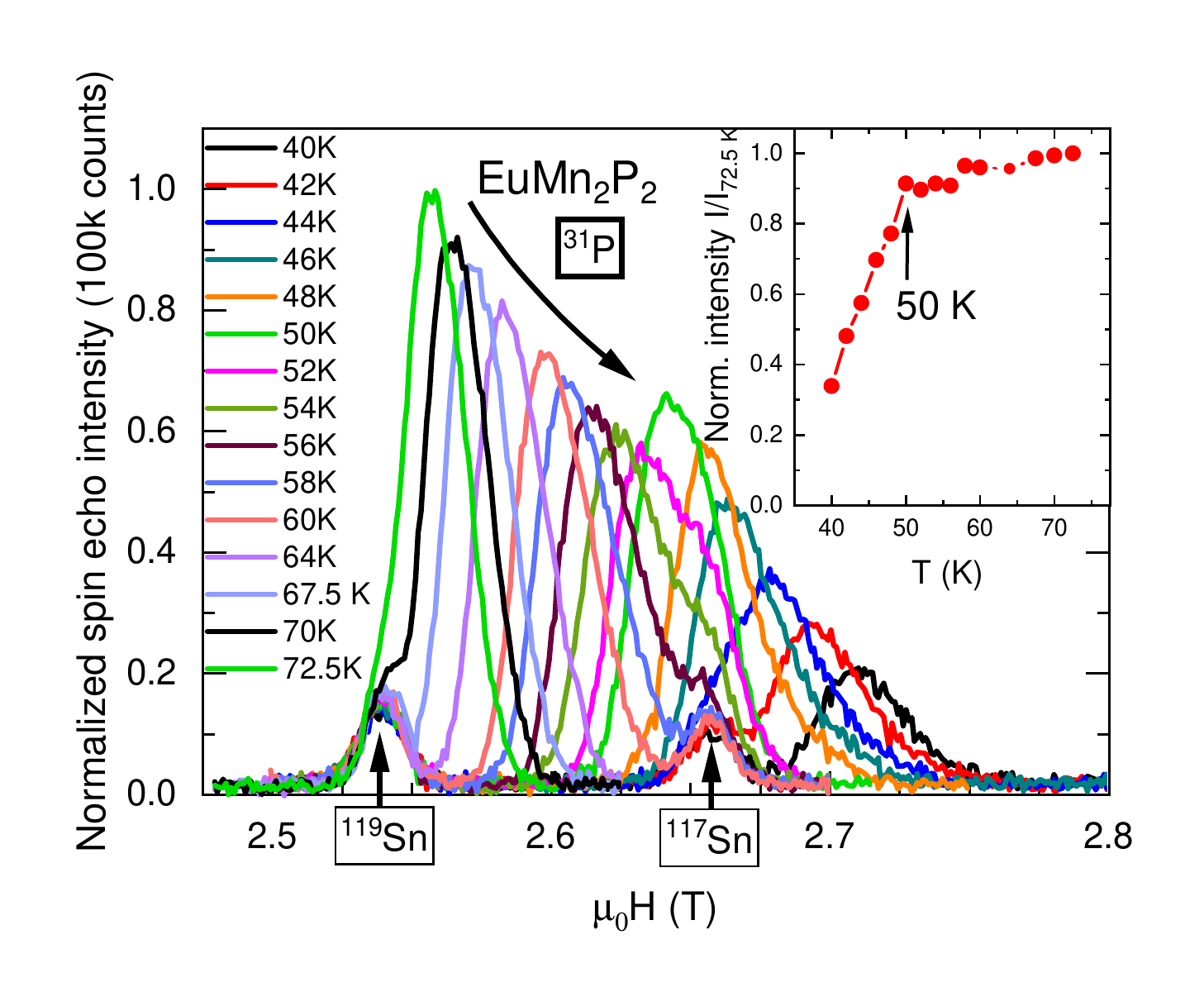}
\caption{\label{fig:NMRFig3} $^{31}$P field sweep NMR spectra taken with an increased number of scans in a narrow temperature window between 72.5 K and 40 K. The spectra were normalized to the maximum signal of the $^{31}$P NMR line at 72.5 K. A strong decrease of the maximum signal is clearly visible. The integration of the signal vs field of the individual $^{31}$P NMR line gives the NMR intensity at each temperature. The NMR intensity of the $^{31}$P sweep NMR line is plotted in the inset as function of temperature (note that this is normalized to the intensity as 72.5 K).}
\end{figure}
Fig.~\ref{fig:NMRFig3} shows the $^{31}$P NMR line always recorded with the same number of scans at various temperatures between 72.5~K and 40~K. Even from these plots, it becomes clear that the line is strongly suppressed towards low temperatures. The NMR intensity is given by the -in field- integration over the NMR line (Note that we corrected for the intensities of the $^{119,117}$Sn lines.) and it is plotted as function of temperature in the inset of the figure. It becomes rather clear that we lose 2/3 of the intensity while going down from 72.5\,K to 40\,K and that the decay sets in at about 50\,K as shown in the inset of Fig.~\ref{fig:NMRFig3}. The full width half maximum (FWHM) of the $^{31}$P NMR line as a function of temperature is shown in Fig.~\ref{fig:NMR_appendix} in the SI \cite{Krebber2025_SI}. 

\subsection{X-ray magnetic circular and linear dicroism}
\begin{figure}
    \centering
    \includegraphics[width=\columnwidth]{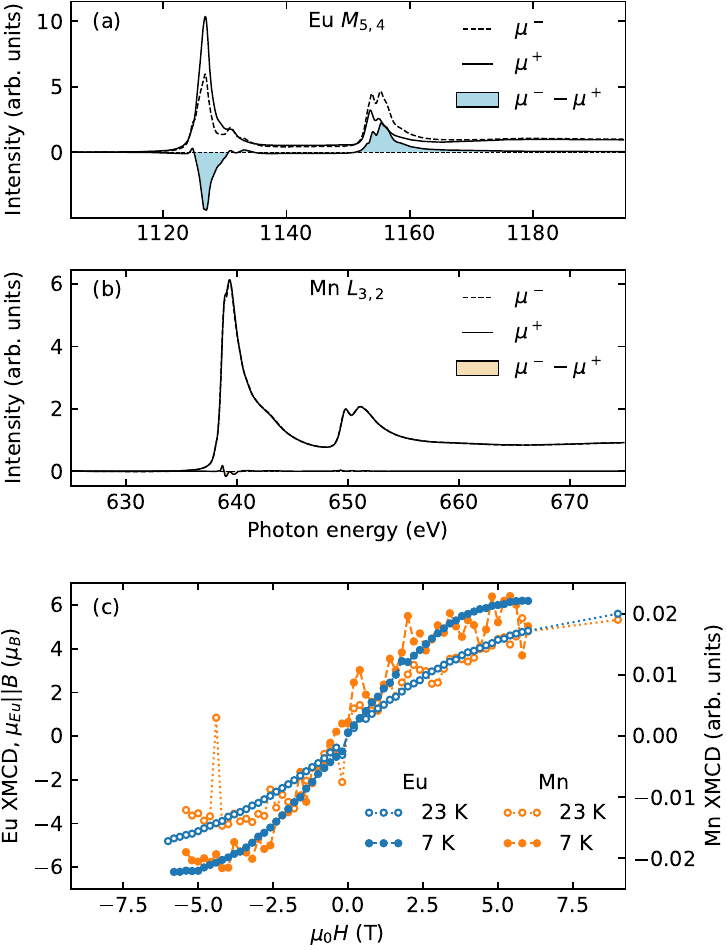}
    \caption{(a) XMCD spectra taken at $T=23\,$K at the Eu $M_{5,4}$ edges, (b) the Mn $L_{3,2}$ edges, and (c) the magnetic field dependence of the XMCD signal at both edges for $T=23\,$K ($T>T_3$) and $T=7\,$K ($T<T_3$).}
    \label{fig:xmcd}
\end{figure}

Having established the existence of weak order which is of magnetic origin through ESR, HC and NMR, we now attempt to gain further information about the transitions at $T_{1,2}$. 
For the element specific magnetic characterization we measured the circular and linear magnetic dichroism in the Mn $L_{3,2}$ and Eu $M_{5,4}$ X-ray absorption spectra as a function of applied magnetic field and temperature. Fig.~\ref{fig:xmcd} shows the X-ray magnetic circular dichroism (XMCD) for both edges at $T=23\,$K and an applied field of $\mu_0H = 9\,$T. In XMCD, the magnetic field is applied along the X-ray beam, H $\parallel c$, and the size of the XMCD signal is proportional to the component of the magnetic moment aligned along the field. Using the XMCD sum rules, we find an aligned magnetic moment of $\mu_{\rm Eu} = \mu_{\rm S} = 5.6\,\mu_B$ at $9\,$T, in agreement with the total magnetic moment per f.u. observed in the bulk magnetization measurements at this temperature. The sum rule analysis yields a pure spin moment with no orbital contribution as expected for Eu$^{2+}$ ions. For the Mn ions, we find only a very small XMCD signal at $T=23\,$K and $\mu_0H = 9\,$T. A precise application of the XMCD sum rules for Mn is usually prohibited by the too small separation between the Mn $L_3$ and $L_2$ absorption edges. However, we can still use the sum rules to estimate an upper limit for the total Mn magnetic moment aligned by the applied field and the spin vs orbital contribution to this moment. We find a very small, pure spin moment without a detectable orbital contribution: $\mu_{\rm Mn} = \mu_{\rm S} < 0.03\,\mu_B$.

In Fig.~\ref{fig:xmcd} (c), we show the field dependence of the Eu and Mn XMCD signal below and above $T_3$. The Eu XMCD signal is again in good agreement with the bulk $M(H)$ measurements. The Mn XMCD signal follows the field dependence observed for Eu within the error bars set by the very small Mn XMCD signal, both in the Eu AFM and paramagnetic phase. The observation of an only very small, pure spin Mn magnetic moment following closely the Eu magnetization could be compatible with an induced magnetic moment at the Mn sites due to the large Eu magnetic moments. 

\begin{figure}
    \centering
    \includegraphics[width=\columnwidth]{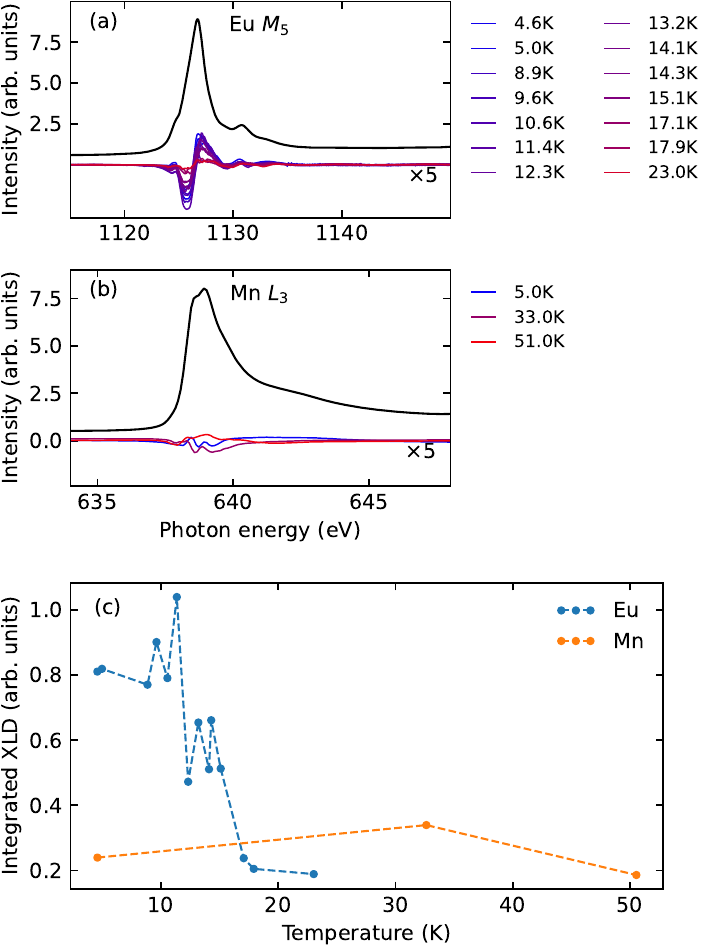}
    \caption{XMLD spectra taken at normal incidence and zero field for Eu (a), at $\mu_0 = 0.4\,$T applied along [100] for Mn (b) as well as the temperature dependence of both the Eu and Mn XMLD signals (c).}
    \label{fig:xmld}
\end{figure}

In Fig.~\ref{fig:xmld}, we present the X-ray magnetic linear dichroism (XMLD) at the Eu $M_{5,4}$ and Mn $L_{3,2}$ edges as a function of temperature for normal incidence, i.e. the incident X-ray beam parallel to [001]. 

In this geometry, the XMLD signal is the difference in the absorption for two orthogonal linear X-ray polarizations in the $a-a$ plane, one along $[100]$ and one perpendicular to it. It is sensitive to the projection of staggered magnetic moments along the two polarization directions, and therefore suitable to detect FM and AFM order alike. 
In normal incidence geometry it can be used to probe the in-plane component of a potential Mn magnetic order.
We can detect a clear XMLD signal at the Eu M$_5$ edge with a strong temperature dependence between 5~K to 18~K, in agreement with the $T_{3}$ detected by other techniques. In addition to this clear temperature dependence in zero applied field, we further confirmed the magnetic origin of the detected signal by applying a magnetic field at $T=5\,$K. With an applied magnetic field of $\mu_0H = 0.4\,$T along [100] we observed an enhancement of the XMLD signal, in agreement with the expectation of going towards a single domain AFM state with moments parallel to [100]. Also in agreement with expectation, applying 9~T along [001] and aligning the Eu moments along this direction suppressed the XMLD signal.

In contrast to Eu, no clear Mn XMLD signal was found within the detection limit of the experiment, even after application of $0.4\,$T along [001]. This was true both below $T_3$ as well as above $T_3$ at $T=33~\rm K$ and $T=51~\rm K$.

\subsection{Resistivity}

\begin{figure}
\centering
 \includegraphics[width=0.85\columnwidth]{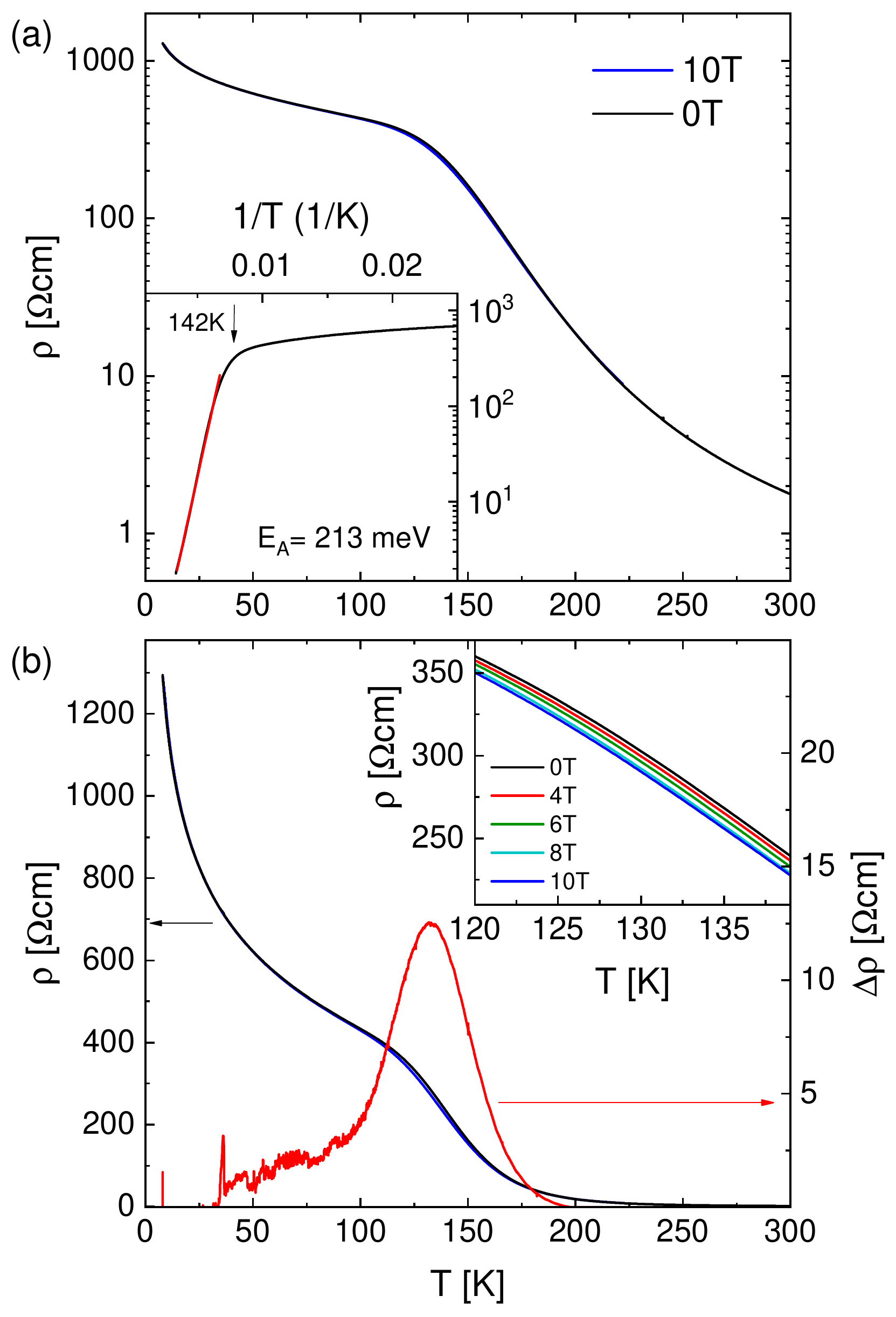}
 \caption{\label{fig:Resistivity} (a) Temperature dependent resistivity with an AC current applied in the $a-a$ plane of EuMn$_2$P$_2$ on a logarithmic scale measured at 0 and 10\,T. The inset shows the Arrhenius plot with a calculated activation energy of E$_A$ = 213\,meV. (b) Temperature dependent resistivity for 10\,T (blue curve) and the difference between the 10\,T and 0\,T curve (red curve).}
 \end{figure}

Fig.~\ref{fig:Resistivity}(a) depicts the resistivity measured with an AC current applied in the $a-a$ plane of EuMn$_2$P$_2$ on a logarithmic scale versus temperature. At room temperature the sample is well conducting with a resistivity of $\rho(T=300$\,K) $\approx$ 2\,$\Omega$cm. Upon cooling down the resistivity increases exponentially as illustrated in the Arrhenius plot in the inset. For high temperatures, a linear fit provides an excellent description of the data, yielding an activation energy of E$_A$ = 213\,meV, in agreement with \cite{Berry2023}. However, at T=142\,K the slope changes, resulting in the emergence of a shoulder-like feature, which appears to be sample dependent, Fig.~\ref{fig:Resistivity_2}. In the same temperature range short-range magnetic correlations of Mn were observed \cite{Berry2023}. At lower temperatures, the resistivity further increases, indicating a continuing, albeit now very weak semiconducting behavior. The blue curve was obtained in an external field of $\mu_0H$ = 10\,T applied along the c-direction. As illustrated by Fig.~\ref{fig:Resistivity}(b), applying a magnetic field along the c-direction does not result in a significant alteration of the overall temperature behavior of the resistivity. This is in stark contrast to analogous compounds such as EuCd$_2$P$_2$ \cite{Wang2021} and EuZn$_2$P$_2$ \cite{Krebber2023}, which exhibit a colossal magnetoresistance behavior at low temperatures. The difference between the $\mu_0H$=10~T and the zero field measurements is plotted in red, where a small negative magnetoresistance of the order of a few percent between T=200\,K and T=100\,K is observed. The inset presents various magnetic fields between 0 and 10\,T, demonstrating a small but systematic shift of the resistivity within this temperature range with increasing magnetic field. 

\section{Discussion}

The most intriguing property of the observed spin resonance in EuMn$_{2}$P$_{2}$ appears when crossing $T_{\rm M}=47$~K where the temperature dependencies show clear anomalies in the linewidth, the resonance field, and the intensity while the lineshape is preserved. Such anomalies are not reported for isostructural EuZn$_{2}$P$_{2}$ where non-magnetic Zn replaces the Mn sites \cite{Goryunov2014}. 
We therefore suspect that the observed anomalies are related to an Eu$^{2+}$ resonance which is coupled to the Mn spin system. 
Above $T=T_{\rm M}$, the Eu relaxation involves Mn spins which have itinerant character as the states near the Fermi level are dominated by Mn in the (Mn$_{2}$P$_{2}$)$^{2-}$ layers \cite{Berry2023}. This itinerancy is reflected in the ESR spectra as a finite dispersion of the lineshape and in a weak Korringa, linear in $T$ linewidth contribution. The Mn spin system has a characteristic temperature $T_{\rm M}$ where short range Mn order might set in. Approaching $T_{\rm M}$ from above, critical Mn-spin fluctuations lead to a broadening of the Eu line $\propto (T-T_{\rm M})^{-1}$. Upon decreasing the temperature below $T_{\rm M}$, the relaxation towards the Mn spin bath is effectively canceled, leading then to the observed kink in the Eu linewidth and to a suppression of the effective Eu-resonance field because of the evolving internal Mn field.
\\
%{\bf HC:\\}
Given the fact, that the signals at $T_{1}$ and $T_2$ detected in the heat capacity are very small, we attribute these high temperature transitions to the weak ordering of the Mn spins in their triangular environment, which might be hindered due to magnetic frustration. The calculated entropy contribution is $S$ = 0.035 Jmol$^{-1}$K$^{-1}$, which represents only 0.6\% of the expected value of Rln2 for a local Mn ordering.
For fields higher than $\approx 1\,\rm T$, the peak at $T_1$ starts broadening and shifting to slightly higher temperature, which hints to a ferromagnetic component of the weak Mn order. This emerging weak magnetic order at $T_1$ appears to shift isotropically to higher temperatures with applied magnetic field and is followed by a further transition at $T_2$ for which we observe an anisotropic increase, when applying the magnetic field along the two different directions.
In case of a simple ferromagnetic transition, one expects a strong broadening already at low magnetic fields, ruling out purely ferromagnetic ordering of the Mn spins at the $T_2$ phase transition. We neither observe a suppression of the transition at $T_{2}$ in field which would indicate simple AFM order. Instead, the peak at $T_{2}$ is stabilized in field which hints to a scenario where frustration of the Mn-moments situated on a triangular lattice in this crystal structure might play a role. This order might be suppressed based on anisotropic magnetic interactions due to the layered Mn system. Furthermore, the observed magnetic field dependencies of the anomalies rule out the possibility of purely structural origin of the transitions at $T_1$ and $T_2$.\\
%{\bf Susceptibility:\\}
Already below $\approx$ 130 K, we observe a minor deviation of $\chi^{-1}(T)$ from the fit according to the Curie-Weiss law, which is consistent with \cite{Berry2023} where the development of Mn magnetic fluctuations is described in a similar temperature range.
In $\chi(T)$, a minor change of slope is observed around the $T_1$ and $T_2$ transitions only. Nevertheless, a comparison of the EuMn$_2$P$_2$ data with those reported earlier for the non-magnetic compounds, (Sr,Ca)Mn$_2$P$_2$ \cite{Sangeetha2021}, would suggest that any contributions of the Mn atoms to the magnetic susceptibility would be of the order of magnitude of 10$^{-10}$m$^3/\rm mol$, which in this temperature range is four order of magnitudes lower compared to the paramagnetic signal from the Eu-moments. This makes it almost impossible to discern weak 3d-based magnetic signatures on the background of an Eu$^{2+}$ system. The minor contribution to the magnetic susceptibility might be attributed to spin fluctuations, which stem from the quasi-two-dimensional nature of the Mn spin layers, along with a possible contribution from magnetic frustration. A similar behavior was observed in CaMn$_2$P$_2$ and SrMn$_2$P$_2$, where the magnetic susceptibility increases with temperature, showing a broad maximum unlike most local moment antiferromagnets \cite{Sangeetha2021}. 
\\
The magnitude of the hyperfine coupling constant as determined by NMR is common for 4f local moment metals \cite{Bruning2008, Sarkar2012}. It should be mentioned that the utilized model is only valid down to approximately 50\,K. Below that temperature, the breakdown of the linearity in $\mathcal{K}$ vs $\chi$ (inset of Fig.~\ref{fig:NMRFig2}) indicates a change of the magnetic correlations in the system (where $A_\mathrm{hf}$ becomes \textit{T}-dependent). This might be taken as a fingerprint of the hidden 3d-Mn-based magnetism. There is also a discontinuous behavior of the linewidth of the  $^{31}$P NMR line, which also might indicate the presence of small Mn based magnetism. Another hint towards the presence of weakly Mn-based magnetism embedded in the predominant local moment 4f Eu based paramagnetism comes from the NMR intensity itself.\\
Based on this, we speculate that around 50\,K the Mn based magnetism is altered, probably undergoes either long range or short ranger order, and the associated critical fluctuations are responsible for the signal wipe out effect found in NMR. From the linewidth change below 100\,K and the determined hyperfine coupling constant, we estimated an upper limit of the moment variation of about 0.06$\mu_B$. This indicates a rather small effect which is not visible in the NMR shift or bulk susceptibility but in the magnetic relaxation process (leading to the wipe out) and in the line broadening. For EuMn$_2$P$_2$ there is only one $^{31}$P NMR study on polycrystalline samples above 100\,K \cite{Berry2023}. 
The data determined there are in reasonable agreement with our data although they were determined in a larger magnetic field (frequency). Absolute shift values and the hyperfine coupling constant (-4\,kOe$/\mu_{\text{B}}$) are close to ours (-5.11\,kOe$/\mu_{\text{B}}$). In contrast, the pure Mn-based system SrMn$_2$P$_2$ has very small shift values, which is due to the weak Mn magnetism \cite{Sangeetha2021}. However, the hyperfine interaction constant determined there (-5.23\,kOe$/\mu_{\text{B}}$) is comparable to the constant determined by us in EuMn$_2$P$_2$ (-5.11\,kOe$/\mu_{\text{B}}$). 

The results of ESR, heat capacity, magnetization, and NMR convincingly show the presence of weak Mn order below 51\,K. Using XMCD, we find a very small, pure spin moment of $\mu_{\rm Mn} = \mu_{\rm S} < 0.03\,\mu_B$ without a detectable orbital contribution, which follows closely the Eu magnetization in field. This could be compatible with an induced magnetic moment at the Mn sites due to the large Eu magnetic moments. No evidence of Mn order independent of Eu can be found in the XMCD data, within the detection limits of the measurement. Also, no clear Mn XMLD signal was found below the $T_{1,2}$ transitions or in the AFM ordered phase of Eu.\\

One possibility why no clear magnetic ordering of Mn was found in XMLD could be, that Mn orders completely out of plane and 0.4\,T cannot realign it in the plane. In this case, Mn would not give an XMLD signal in the used geometry. We should also keep in mind the potential influence of the surface in the discussed X-ray magnetic dichroism experiments. In that regard, there are examples, like GdRh$_2$Si$_2$ or EuRh$_2$Si$_2$, where surface and bulk show different ordering temperatures \cite{Guettler2016}, \cite{Chikina2014} or others, like HoIr$_2$Si$_2$, where despite AFM order of the bulk \cite{Kliemt2018}, no magnetic order of the surface could be detected (unpublished data).\\
The magnetic transitions of the material are also not visible in the electrical transport data. Interestingly, the resistivity exhibits a change of slope at 142\,K which indicates a change of the band structure near the  temperature where in the inverse magnetic susceptibility a slight deviation from the Curie-Weiss behavior is observed which is probably connected to the onset of magnetic fluctuations of Mn.\\

\section{Summary}
We investigated the electron spin resonance of \emp\, and, besides the antiferromagnetic transition of Eu at 18\,K, we found an additional feature at 47\,K which is of magnetic origin.
The ESR temperature dependencies show clear anomalies in the linewidth, the resonance field, and the intensity, while the lineshape is preserved when crossing $T_{M}=47$\,K.
Using heat capacity, we could resolve two small subsequent transitions $T_1$ and $T_2$ in the same temperature range. The transition at $T_1=51~\rm K$ is isotropic for field along the two crystallographic main symmetry directions, shifts to higher temperatures and broadens in field which hints to its possibly ferromagnetic nature. Given the small size of the signal, we attribute this transition to a possible order of the Mn spins. 
The transition at $T_2=48.5\,\rm K$ is anisotropic, gets enhanced and shifts to higher $T$ in field. This particular behavior at the transition is probably caused by magnetic frustration favored by the trigonal arrangement of the Mn sublattice.
Using NMR, we can give an upper limit for the small moment of Mn of only $0.06\,\mu_B$. This small magnetic contribution emerging at $T_{1,2}$ on a large paramagnetic background of the Eu spins is in accordance with the electrical resistivity of \emp\, showing no sign of the putative ordering of Mn. 
While ESR and NMR are consistent with an evolving internal field of Mn at $T_1=51\,\rm K$, no direct evidence for Mn order is detected via XMCD or XMLD in the same temperature range which might be due to the small size of the magnetic moment, with a potential $c$-component or frustration. 
Our susceptibility and electrical transport data are consistent with the conclusion of a possible onset of Mn fluctuations below $\approx 150\,\rm K$ which was drawn in a previous report \cite{Berry2023}. 
From density functional theory calculations presented in Ref.~\cite{Berry2023} it was concluded that if Mn order occurs, then it should be AFM. This is not supported by our heat capacity data, which rather hints to a FM nature of the Mn order at $T_1$. Further theoretical studies will be necessary in the future in order to fully understand the entire scenario in the EuMn$_2$P$_2$ system.\\

\section{Acknowledgments}
We thank T. Förster for technical support. We thank C.Geibel, R. Valent{\'i}, A. B\"ohmer and H.-A. Krug von Nidda for valuable discussions and acknowledge funding by the Deutsche Forschungsgemeinschaft (DFG, German Research Foundation) via the SFB/TRR 288 (422213477, project A03, A10, B02).   

\bibliographystyle{apsrev4-2}
\bibliography{EuMn2P2-Bib}

\clearpage

\onecolumngrid
\section{Supplemental Material}
\renewcommand{\thefigure}{S\arabic{figure}}
\setcounter{figure}{0}

\subsection{Heat capacity}
\begin{figure}[H]
\centering
\includegraphics[width=0.5\linewidth]{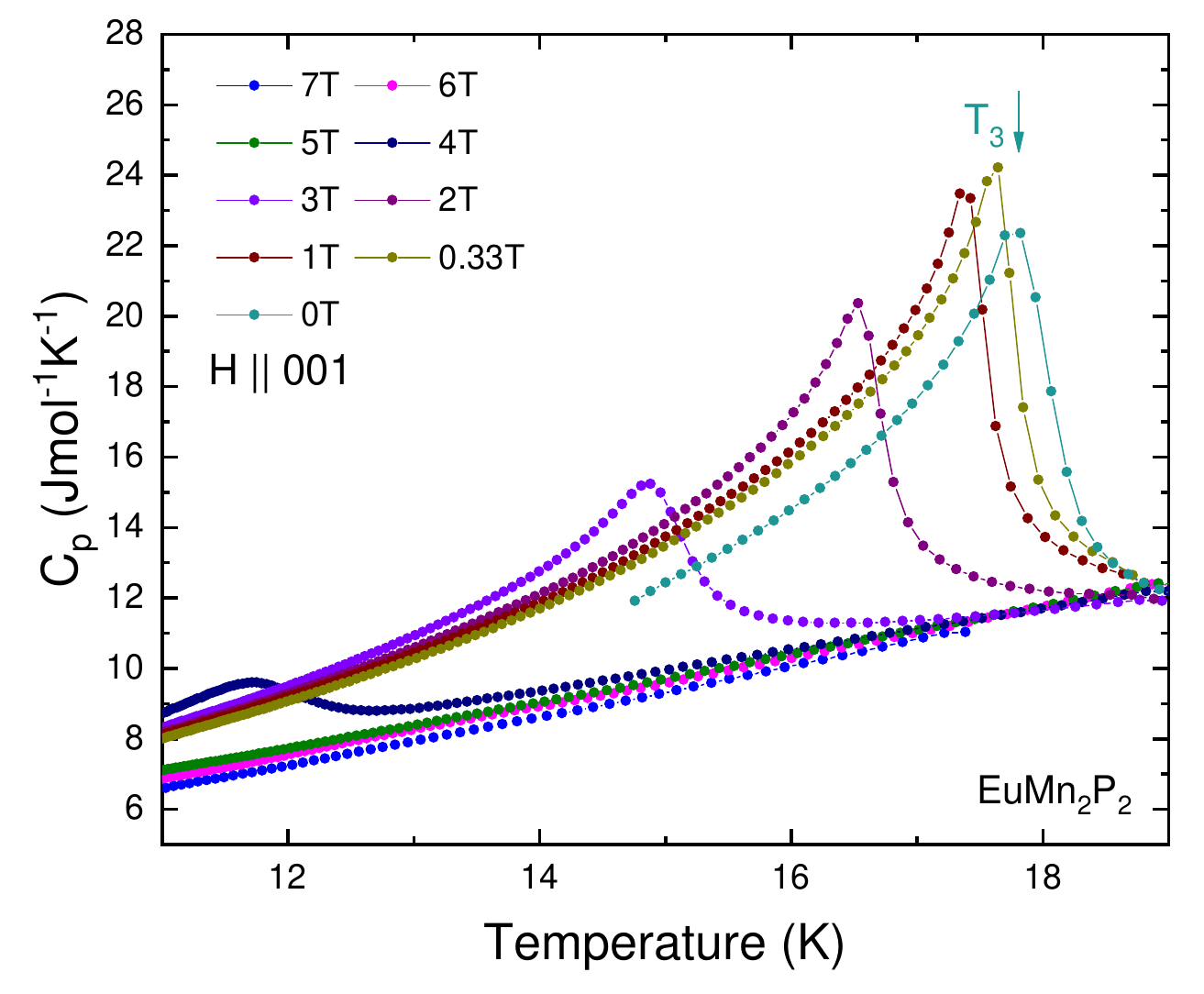}
\caption{\label{fig:AE105_HC_field_Eu} Heat capacity of EuMn$_2$P$_2$ measured in different magnetic fields $H\parallel001$ below 19\,K as a function of temperature. The magnetic transition at $T_3$ is assigned to the AFM ordering of Eu$^{2+}$ and shifts to lower temperatures with increasing field.}
\end{figure}

\begin{figure}[H]
\centering
\includegraphics[width=0.5\linewidth]{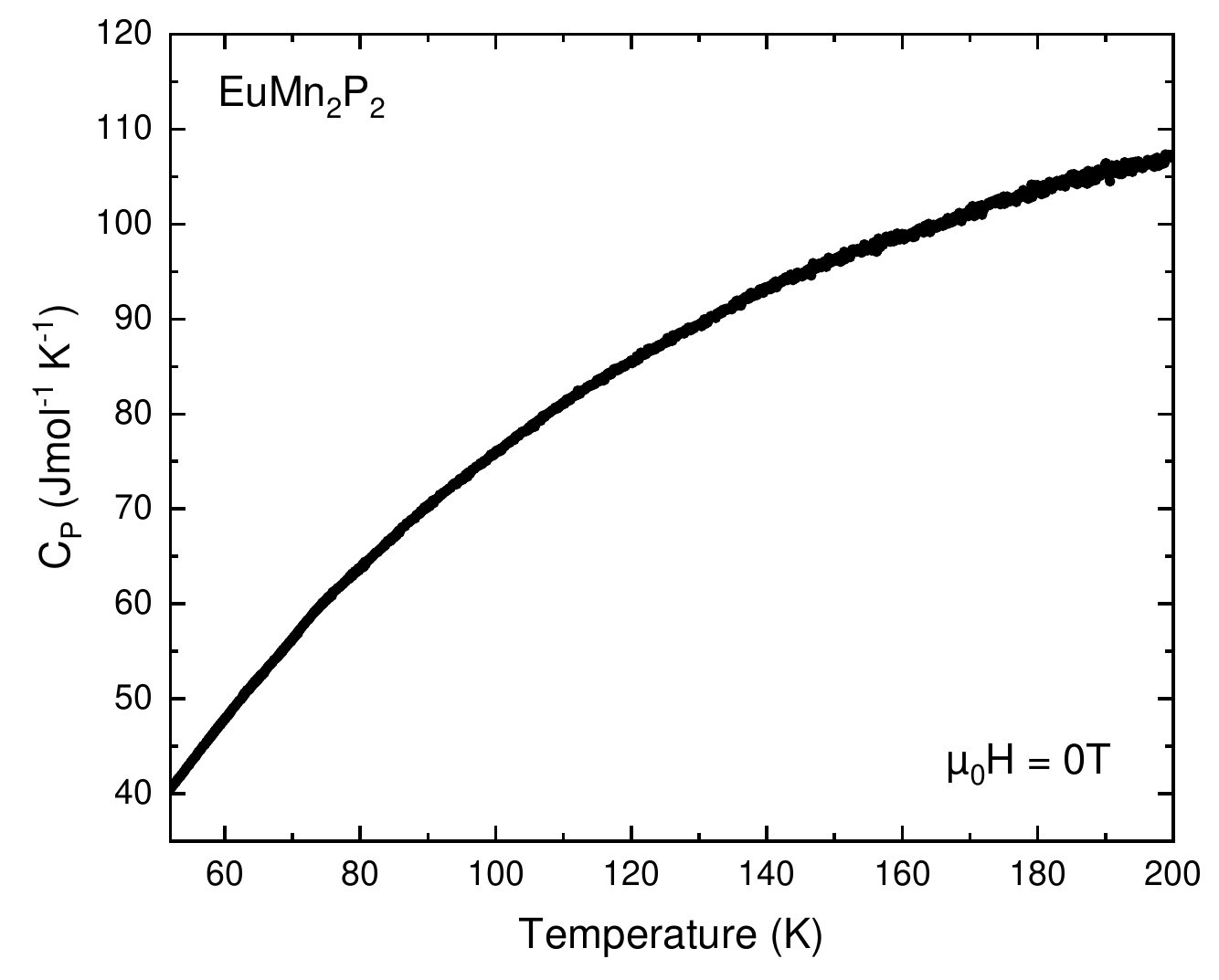}
\caption{\label{fig:HC_hightemp} The zero-field heat capacity of EuMn$_2$P$_2$ as a function of temperature between 52\,K and 200\,K. }
\end{figure}

Fig.~\ref{fig:AE105_HC_field_Eu} shows the field-dependent heat capacity of EuMn$_2$P$_2$ at the T$_3$ transition. At zero field, a peak appears at 18~K which then shifts to lower temperatures when a field is applied along the $c$ direction. The transition comes from the antiferromagnetic ordering of the Eu$^{2+}$ spins. 
Fig.~\ref{fig:HC_hightemp} shows the zero-field heat capacity of EuMn$_2$P$_2$  between 52\,K and 200\,K. No further transitions are observed in this temperature range. 

\subsection{Magnetic susceptibility and inverse susceptibility}
\begin{figure}[H]
\centering
\includegraphics[width=0.45\linewidth]{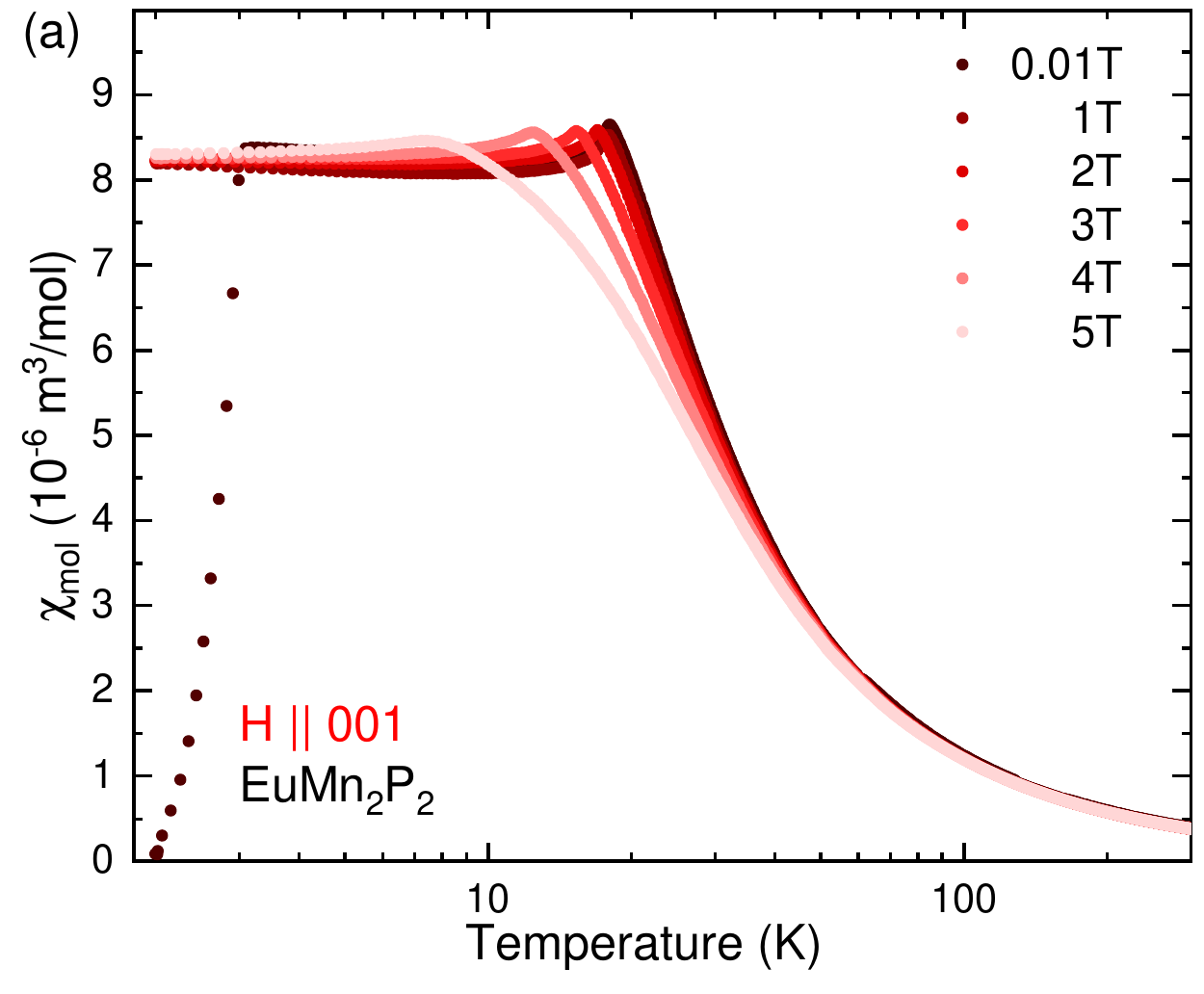}
\includegraphics[width=0.45\linewidth]{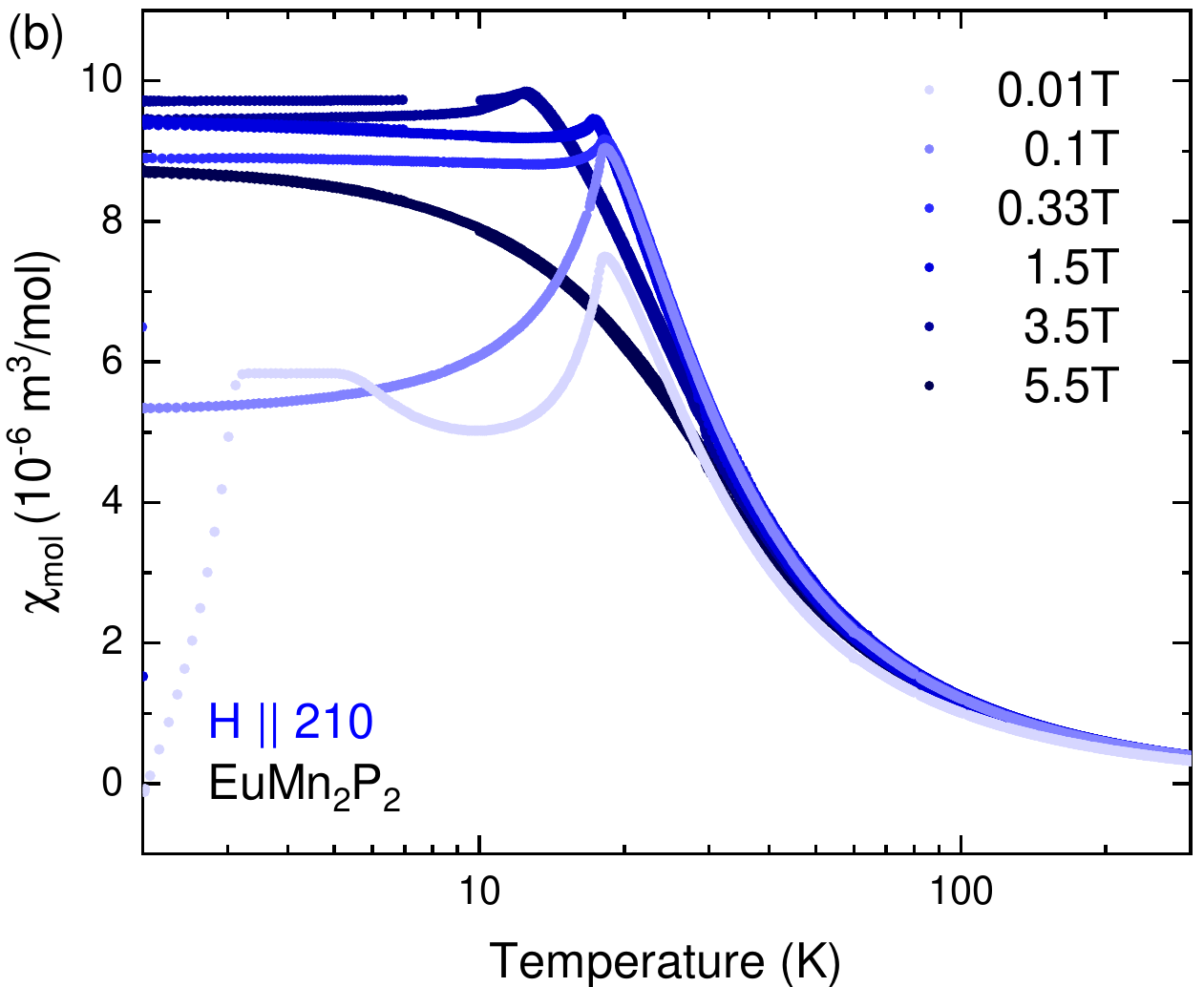}
\caption{\label{fig:AE114_Chi_field001} $\chi_{\rm mol}(T)$ (a) $H\parallel 001$ and (b) $H\parallel 210$: The Eu transition at $T_3$ shifts to lower temperatures with increasing field.}
\end{figure}

The magnetic susceptibility as a function of temperature for the in-plane- and out-of-plane direction is shown in Fig.~\ref{fig:AE114_Chi_field001}. Both directions show a peak at T$_3$ = 18\,K. Upon increasing the magnetic field, the peak shifts to lower temperatures for both directions, which is comparable with the heat capacity data shown in Fig.~\ref{fig:AE105_HC_field_Eu}. The drop at 3.7\,K occurs due to Sn inclusions inside the sample, which become superconducting at low fields. 

\begin{figure}[H]
\centering
\includegraphics[width=0.5\linewidth]{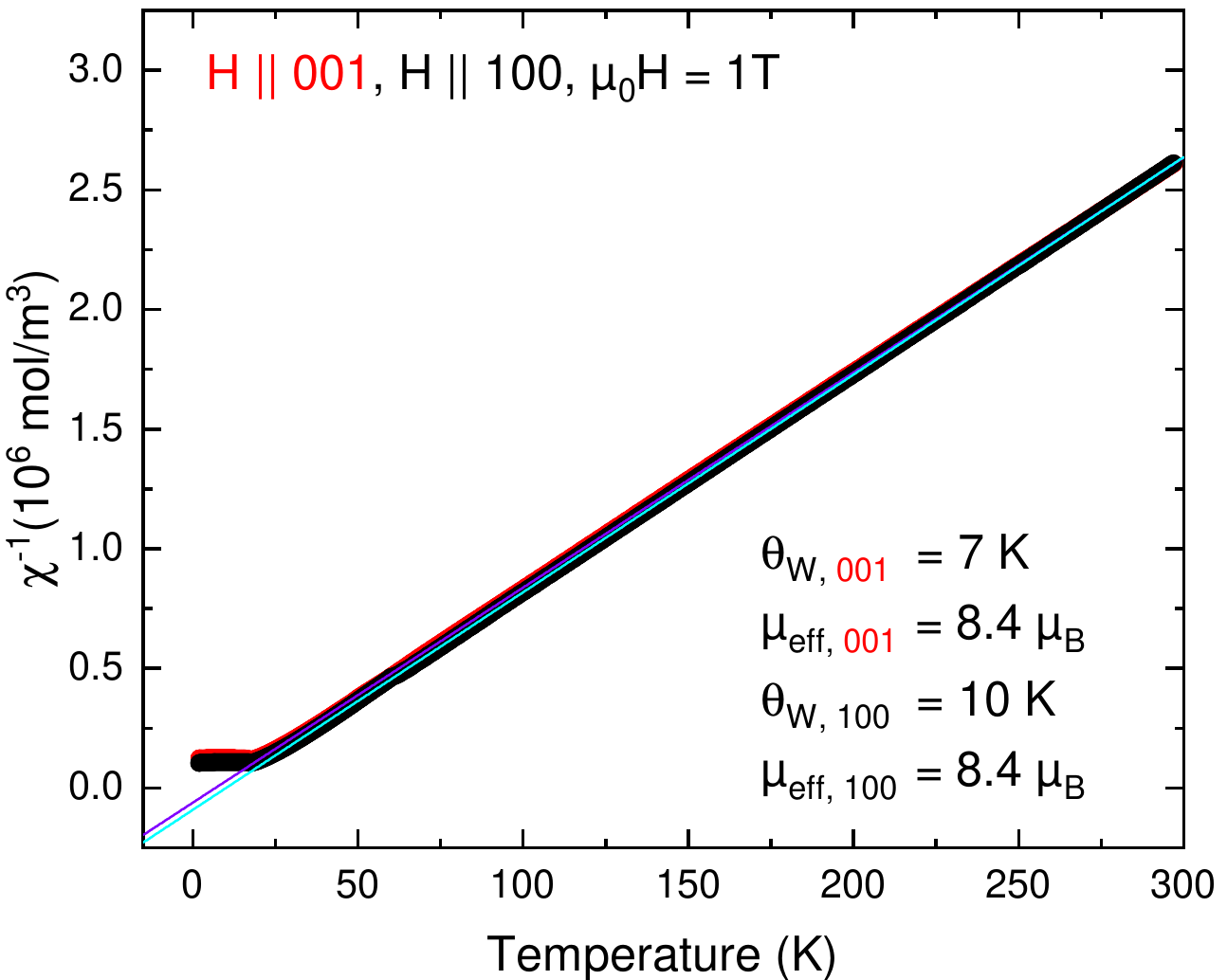}
\caption{\label{fig:AE114_Chi-1}  Inverse magnetic susceptibility $\chi(T)^{-1}$ for $H\parallel 001$ and $H\parallel 100$. The data are fitted using the Curie-Weiss law.}
\end{figure}

The inverse magnetic susceptibility at $\mu_0H=1\,\rm T$ is depicted in Fig.~\ref{fig:AE114_Chi-1} for two field directions. From a Curie-Weiss fit at high temperatures between 150 K and 290 K, the effective moments $\mu_{\rm eff}$ = 8.4 $\mu_B$ were calculated for field applied along the out-of-plane and in-plane directions.

\begin{figure}[H]
\centering
\includegraphics[width=0.5\linewidth]{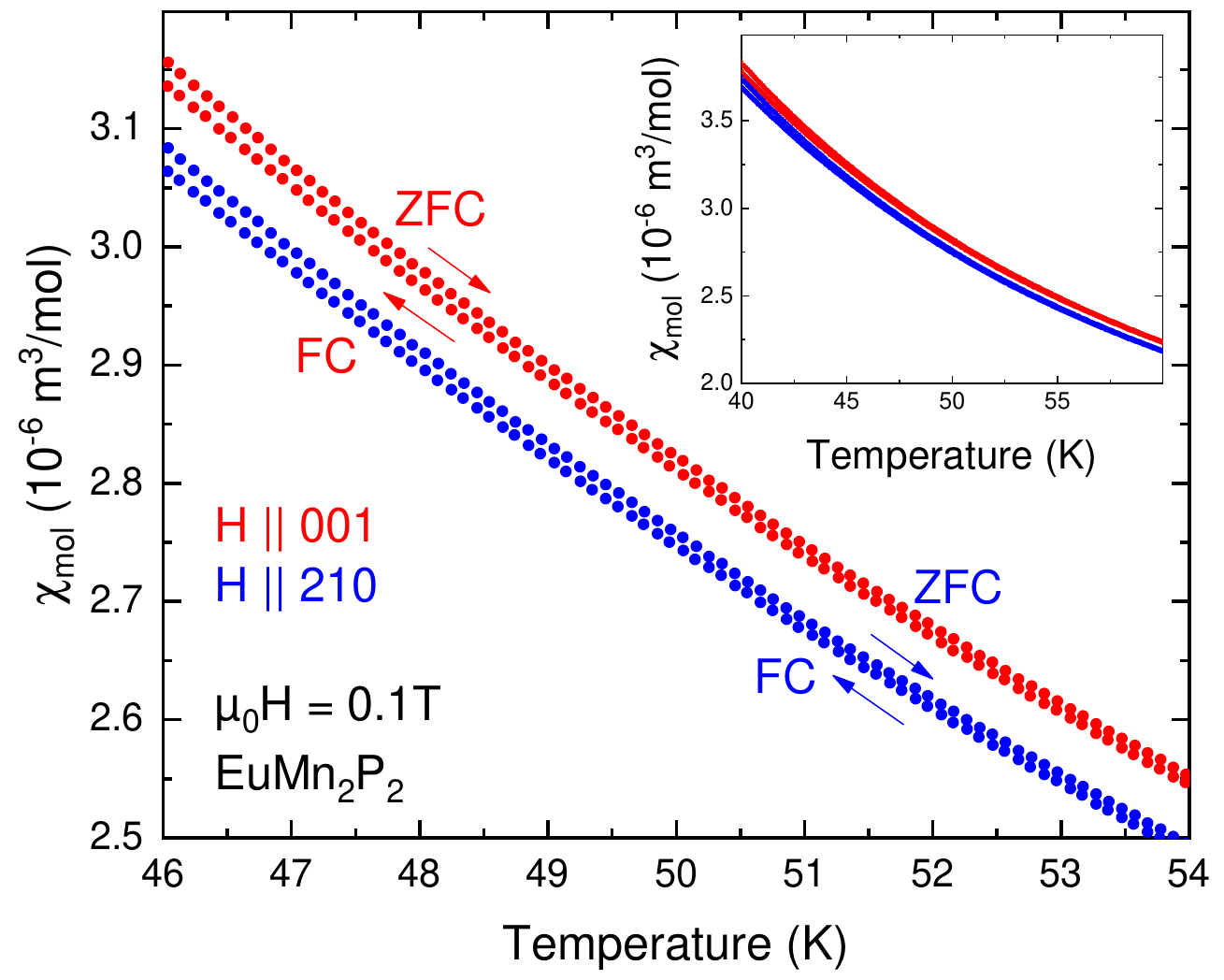}
\caption{\label{fig:AE109_ChivT_FC-ZFC} Comparison of $\chi_{\rm mol}(T)$ with $H\parallel 001$ (red) and $H\parallel 210$ (blue) close to 50 K measured in FC and ZFC mode.}
\end{figure}
In Fig.~\ref{fig:AE109_ChivT_FC-ZFC}, susceptibility data recorded in zero-field cooled and field-cooled mode close to the transitions at $T_{1,2}$ are depicted.
While the transition, attributed to magnetic order of Mn is visible in HC, there is no clear feature in $\chi(T)$. If at all, a slight difference between ZFC and FC curve is visible below $\approx$ 50~K.

\clearpage
\subsection{Field dependence of the moment}

\begin{figure}[H]
\centering
\includegraphics[width=0.44\linewidth]{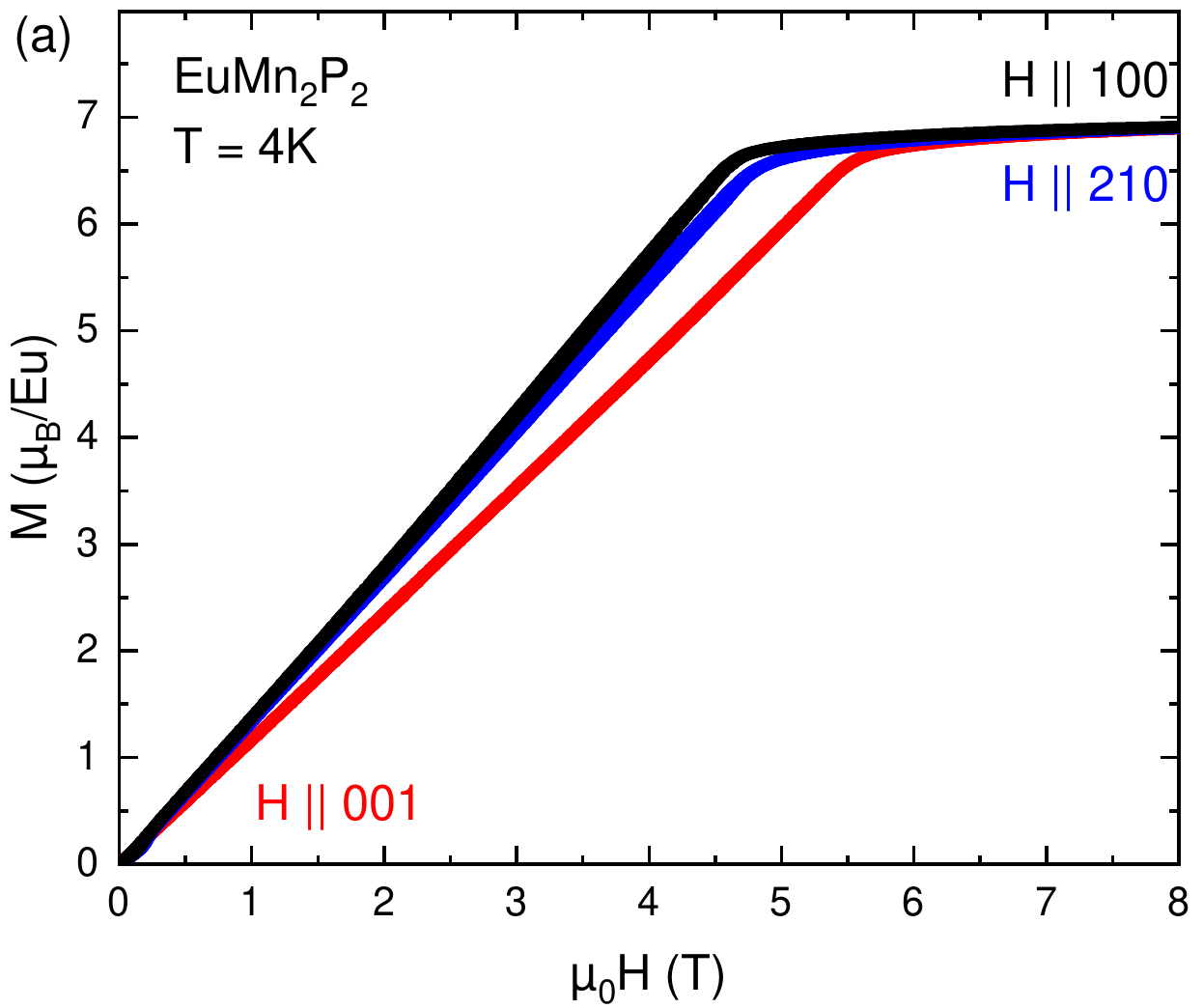}
\includegraphics[width=0.45\linewidth]{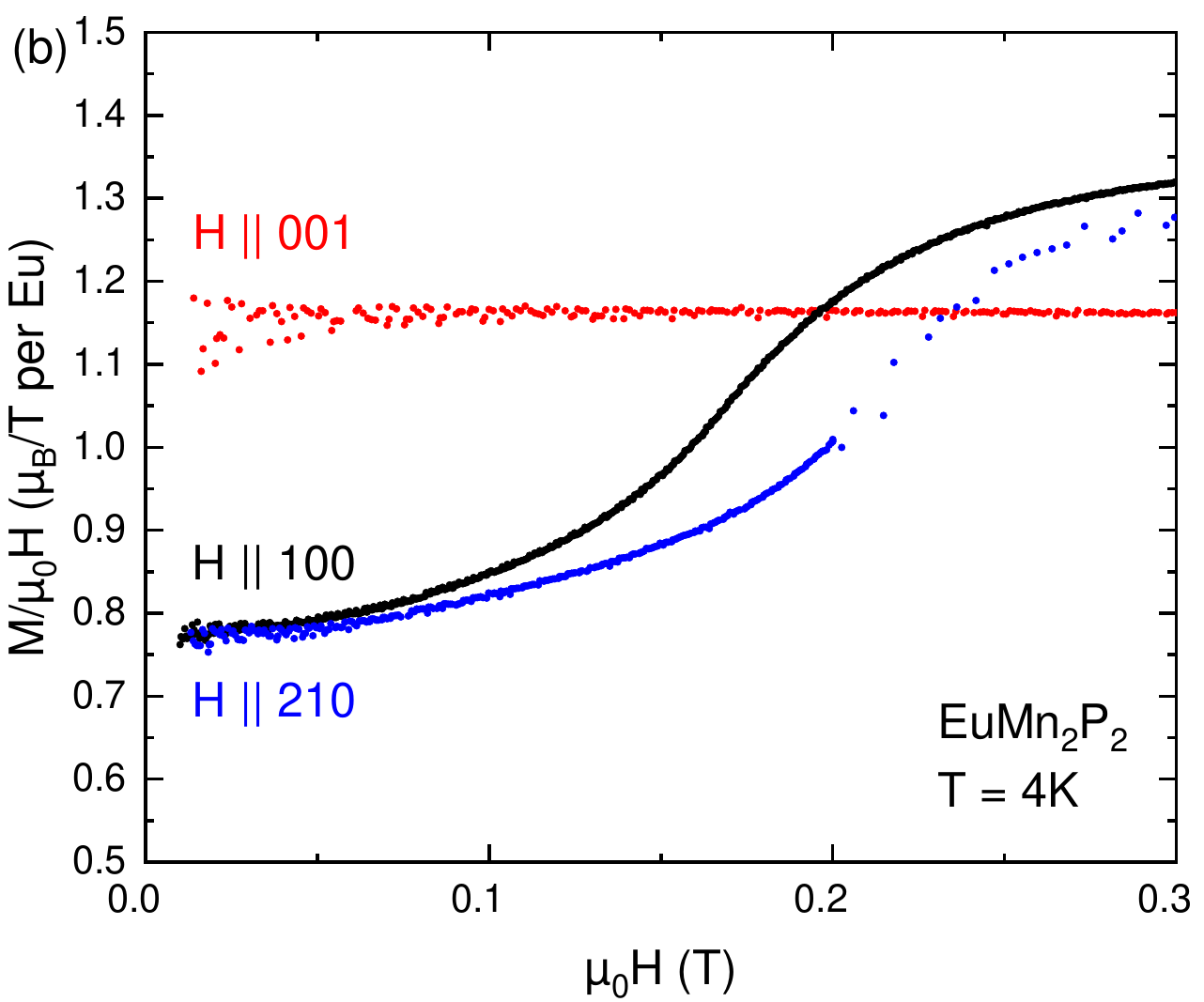}
\includegraphics[width=0.45\linewidth]{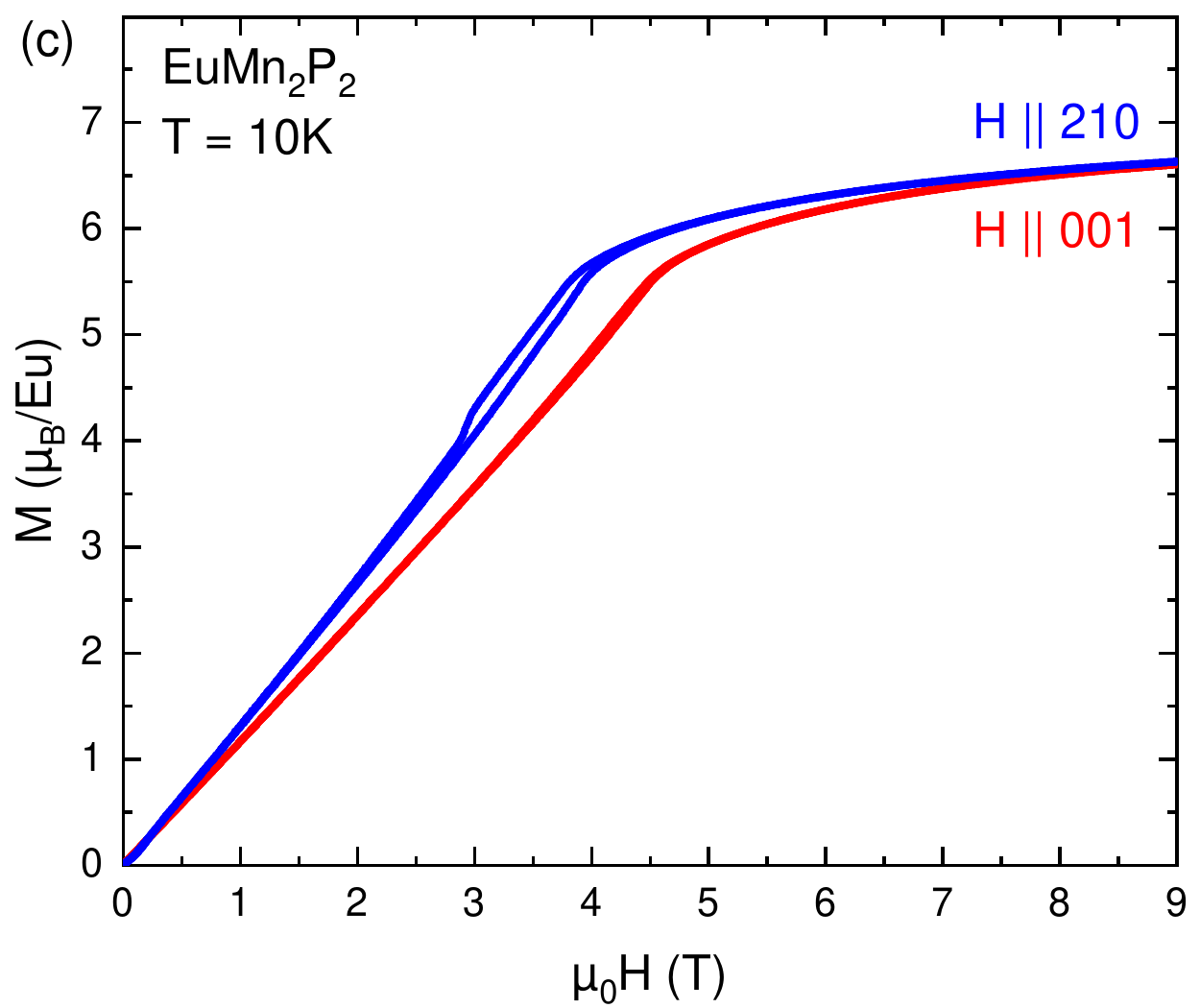}
\caption{\label{fig:AE114_MvH_Vgl4K} Comparison of in-plane $[100]$, $[210]$, and out-of-plane $[001]$ direction (a) Magnetic moment $M(H)$ at 4 K. (b) The low-field magnetic susceptibility $M/H(H)$ versus field at 4~K reveals the in-plane/out-of-plane anisotropy as well as a weak in-plane anisotropy. \label{fig:AE114_MdurchH_4K} (c) \label{fig:MvH_10K.pdf} Magnetic moment $M(H)$ at 10~K.}
\end{figure}

Fig.~\ref{fig:AE114_MvH_Vgl4K}(a)  shows the moment versus field of EuMn$_2$P$_2$ for field along two distinct in-plane directions, $[100]$ and $[210]$, and the out-of-plane direction $[001]$. At 4\,K, the saturated moment reaches $\approx$ 7$\mu_B$. The critical fields are $\mu_0H^{100}_c= 4.7\,\rm T$, and $\mu_0H^{001}_c= 5.5\,\rm T$. This is in contrast to a recent report \cite{Berry2023} where $\mu_0H^{001}_c\approx 4\,\rm T$ was found.

\begin{figure}[H]
\centering
\includegraphics[width=0.45\linewidth]{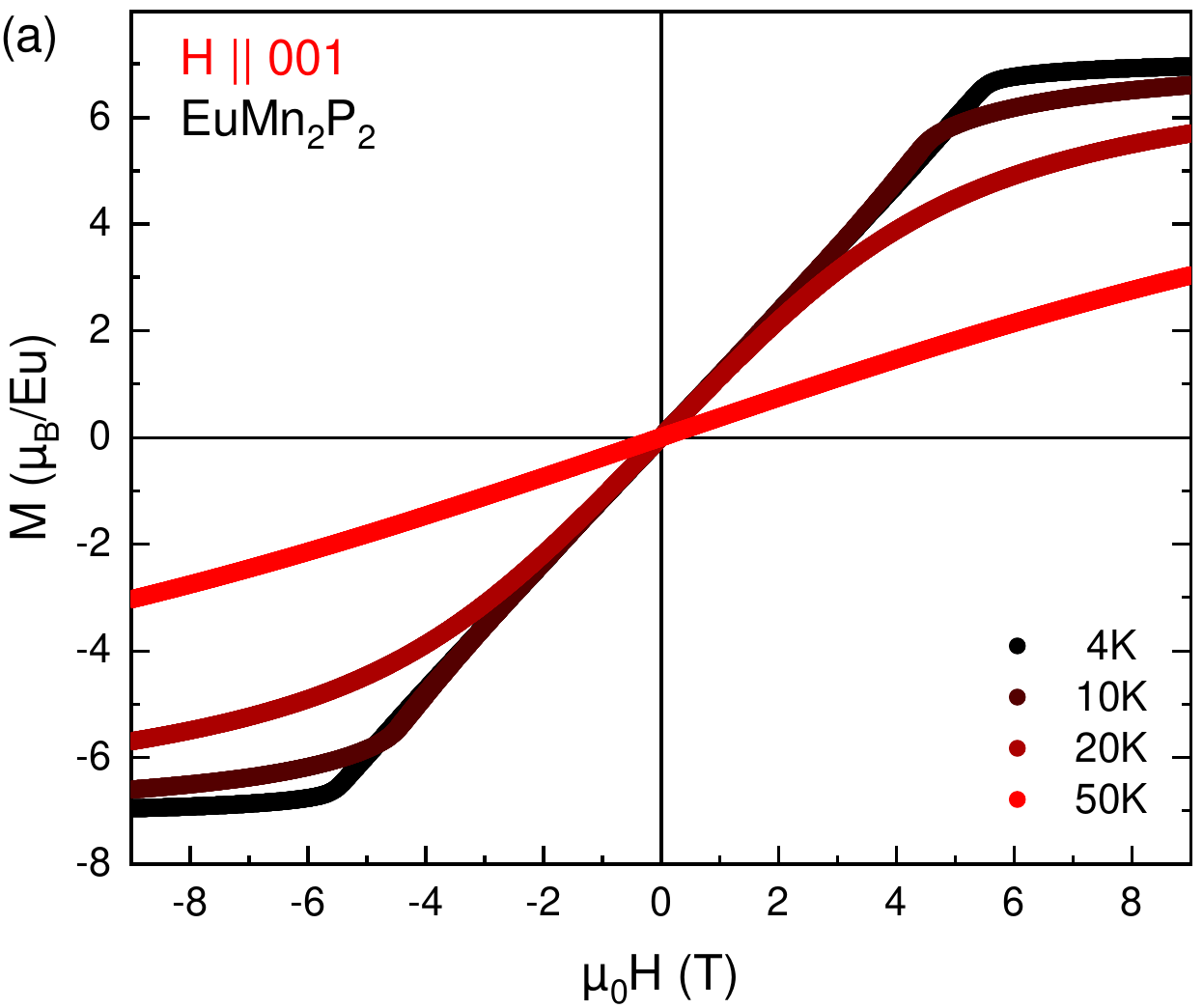}
\includegraphics[width=0.45\linewidth]{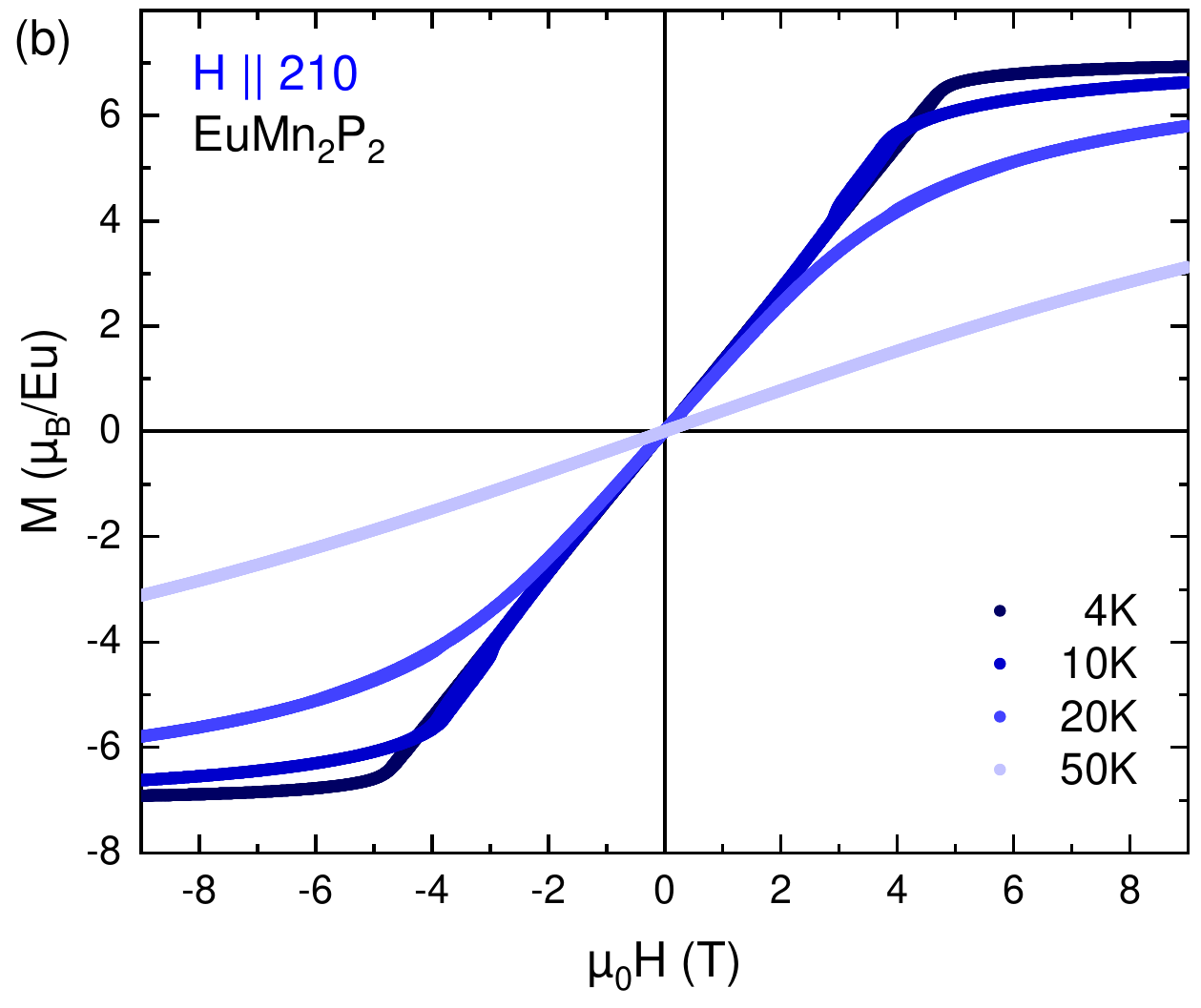}
\caption{\label{fig:AE114_MvH_210} The moment versus field $M(H)$ for (a) $H\parallel 001$ and (b) $H\parallel 210$ is dominated by the strong magnetism of Eu.}
\end{figure}

\begin{figure}[H]
\centering
\includegraphics[width=0.5\linewidth]{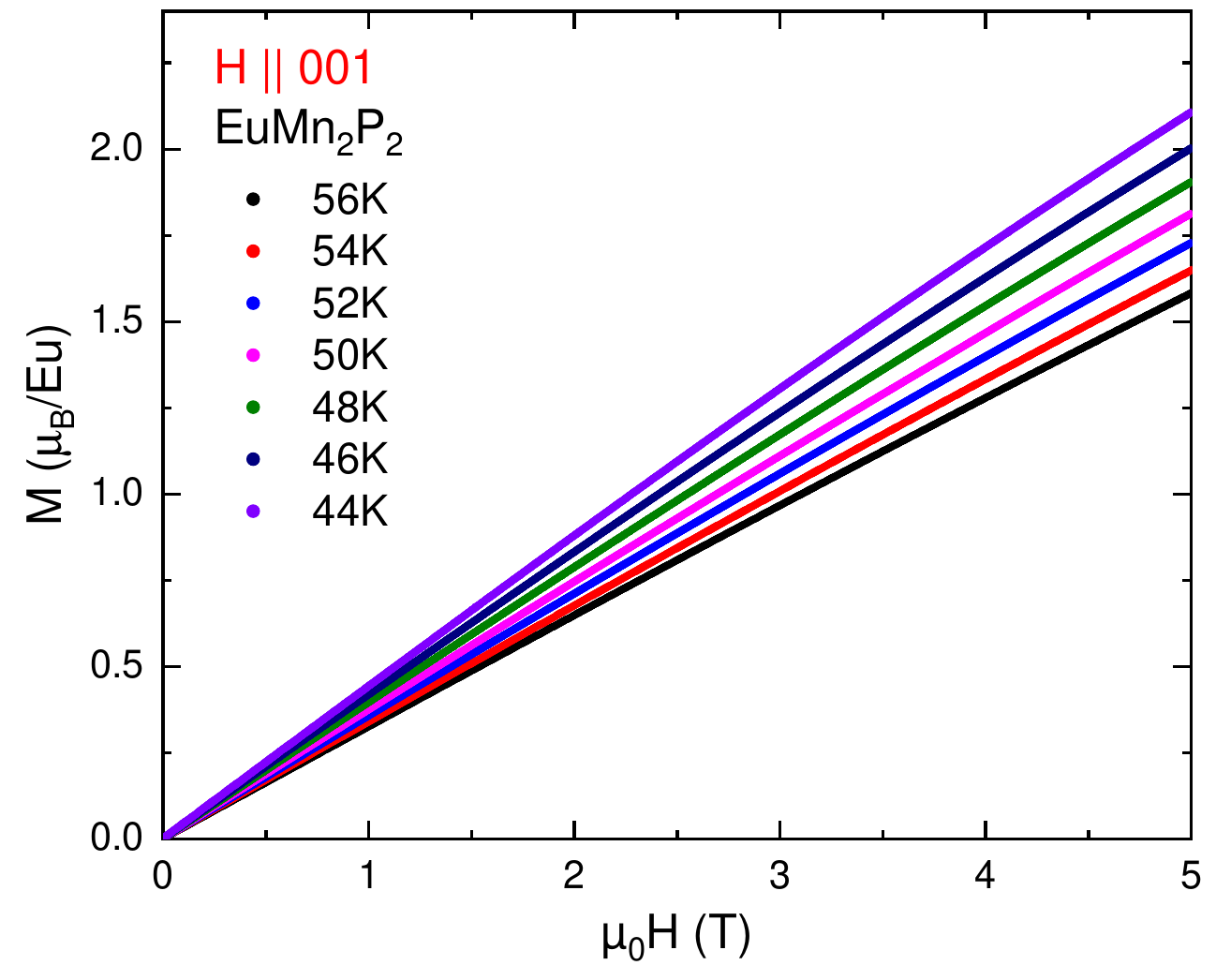}
\caption{\label{fig:AE114_MvH_50K} $M(H)$ at high T close to the putative Mn transition. }
\end{figure}

From Fig.~\ref{fig:AE114_MdurchH_4K}(b) it is apparent that EuMn$_2$P$_2$ shows a very low anisotropy between the in-plane $[100]$ and $[210]$ directions, and the $M/\mu_0H$ data with field aligned along the $[001]$ direction is almost constant for low fields. When applying the field along the $[100]$ direction, the slope starts changing at a field of $\approx$ 0.1\,T which is slightly lower than the field of $\approx$ 0.15\,T where the slope change occurs in case of field along the $[210]$ direction. The constant value for the $[001]$ directions suggest that the magnetic Eu moments are aligned predominantly in the $a-a$ plane, while the particular response for in-plane fields indicates that in the system with the six AFM domains, the magnetic moments are aligned close to the $[100]$ and symmetric equivalent directions. The data clearly show that at 4~K, the alignment of the Eu spins is not changed by applying a field up to $\mu_0H \approx 0.05\,\rm T$. This is in contrast to the related EuZn$_2$P$_2$ \cite{Krebber2023} and EuCd$_2$P$_2$ \cite{Usachov2024} compounds, where the Eu spin arrangement and domain distribution can be changed using much lower fields. At 10~K, a stronger hysteresis between 2 and 4\,T is found for field along the in-plane direction, as shown in Fig.~\ref{fig:MvH_10K.pdf}(c). This result is in contrast to the data presented in Ref.~\cite{Berry2023} where the occurrence of a hysteresis in the same field range was observed for $H\parallel 001$. 
Fig.~\ref{fig:AE114_MvH_210}(a,b) show the comparison of $M(H)$ at different temperatures with field applied along $[210]$ and $[001]$.
We furthermore recorded $M(H)$ data close to the $T_{1,2}$ transitions, Fig.~\ref{fig:AE114_MvH_50K} and found no significant change of slope which is consistent with a very weak magnetic contribution from Mn. 

\clearpage
\subsection{Resistivity}

\begin{figure}[H]
\centering
\includegraphics[width=0.5\linewidth]{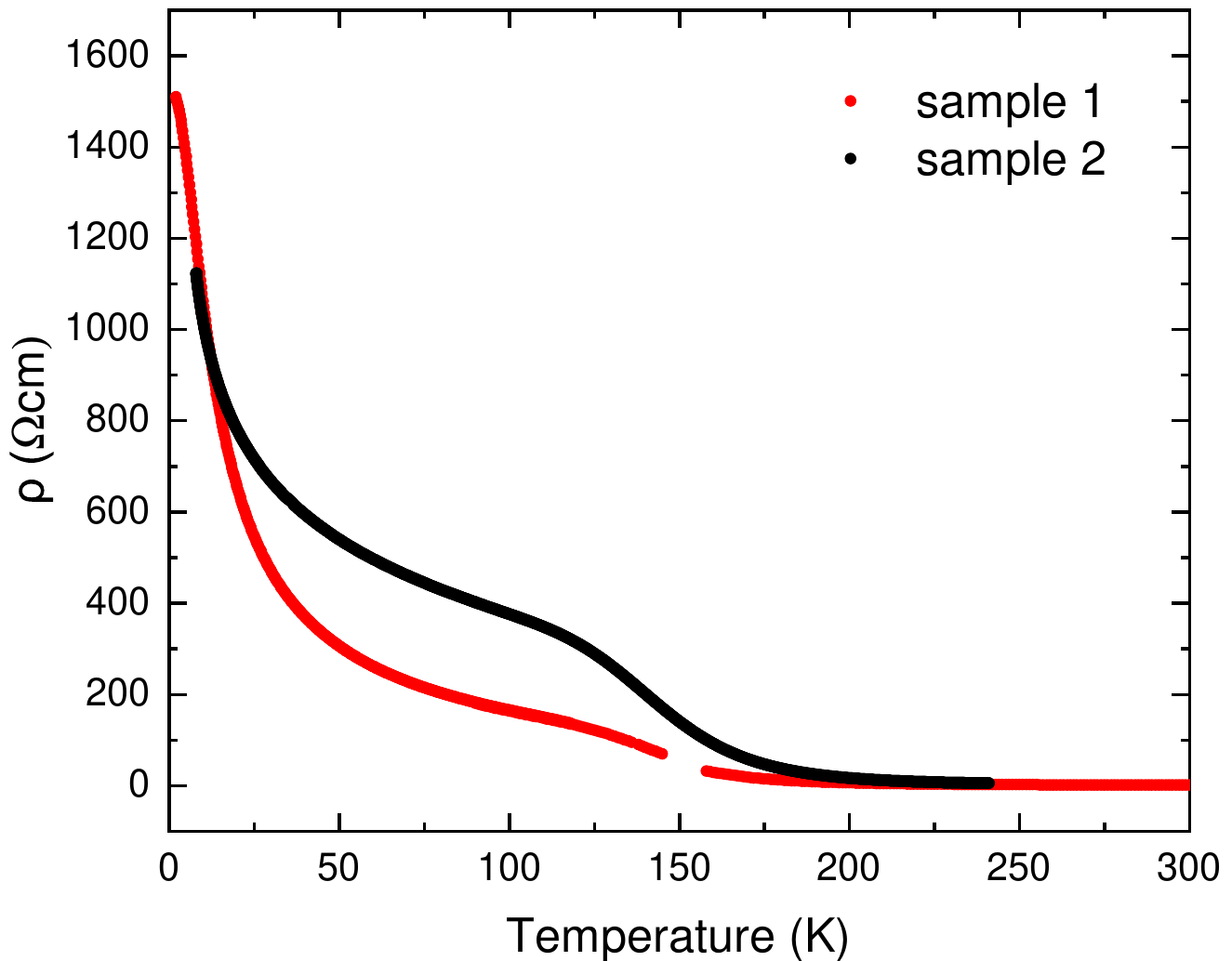}
 \caption{\label{fig:Resistivity_2} Temperature dependence of the resistivity $\rho$ at $\mu_0H$ = 0\,T with the current in the $a-a$ plane for two different single crystals.}
 \end{figure}

Fig.~\ref{fig:Resistivity_2} shows the temperature dependent resistivity for two different EuMn$_2$P$_2$ single crystals, which exhibits a certain sample dependence. At room temperature, both samples are well conducting and upon cooling become increasingly insulating. The resistivity of both crystals shows a broad hump at around $\approx$150\,K, however the slope slightly deviates for both samples.

\clearpage
\subsection{NMR}

\begin{figure}[H]
\centering
\includegraphics[width=0.6\linewidth]{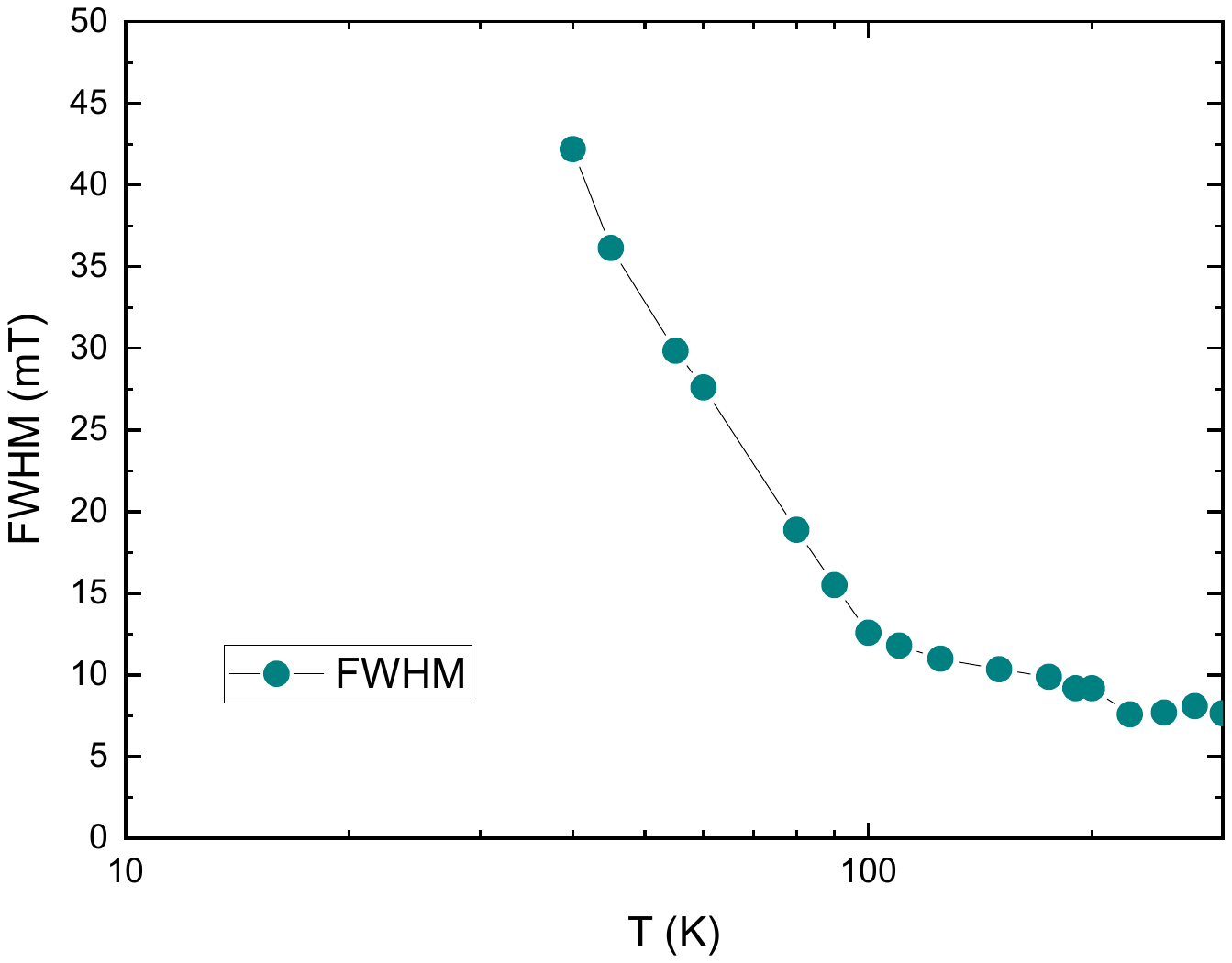}
\caption{\label{fig:NMR_appendix} Full width half maximum (FWHM) of the $^{31}$P NMR line as a function of temperature.}
\end{figure}

Line width FWHF (full width at half maximum) plotted as function of temperature in Fig.~\ref{fig:NMR_appendix}.
It is clearly recognizable that the line width increases monotonically from room temperature to lower temperatures and then shows a significantly stronger increase below $\approx$100\,K. We suspect that this has to do with the onset of critical fluctuations in preparation for the Mn-based order. 

\end{document}